\documentclass[a4paper,fleqn]{cas-dc}

\usepackage[numbers]{natbib}
\usepackage{caption}
\usepackage{amsmath,amssymb,amsfonts}
\usepackage{algorithm,algorithmic}
\usepackage{graphicx}
\usepackage{textcomp}
\usepackage{xcolor}

\usepackage{graphicx}
\usepackage{here}
\usepackage{array}
\usepackage{color,amsmath,here}
\usepackage{amssymb}
\usepackage{enumerate}
\usepackage{ascmac}
\usepackage{tascmac}
\usepackage{fancybox}
\usepackage{float}
\usepackage{amsmath}
\usepackage{mathrsfs}
\usepackage{fancybox}
\usepackage{amsthm,mathtools}

\newtheorem{definition}{Definition}
\newtheorem{assumption}{Assumption}
\newtheorem{prob}{Problem}
\newtheorem{theo}{Theorem}
\newtheorem{rem}{Remark}
\newtheorem{propo}{Proposition}

\newtheorem{lem}{Lemma}

\newcommand{\bm}[1]{{\mbox{\boldmath $#1$}}}

\definecolor{mygreen}{rgb}{0, 0.7, 0}
\definecolor{myyellow}{rgb}{0.7, 0.7, 0}
\definecolor{mypurple}{rgb}{0.42, 0, 0.84}

\def\rank{{\rm rank} \hspace{0.3mm}}

\def\im{{\rm im} \hspace{0.3mm}}
\def\ker{{\rm ker} \hspace{0.3mm}}
\def\tr{{\rm tr} \hspace{0.3mm}}
\def\TAU{\scalebox{0.98}{${\mathcal T}$}}
\def\sprad{{\rm sprad}\hspace{0.3mm}}
\def\EFF{\mathcal F}

\def\hsymbu#1{\smash{\lower1.9ex\hbox{\scalebox{1.0}{$#1$}}}}

\def\tsc#1{\csdef{#1}{\textsc{\lowercase{#1}}\xspace}}
\tsc{WGM}
\tsc{QE}
\tsc{EP}
\tsc{PMS}
\tsc{BEC}
\tsc{DE}

\begin{document}
\let\WriteBookmarks\relax
\def\floatpagepagefraction{1}
\def\textpagefraction{.001}
\shorttitle{Policy Gradient Methods for Designing Dynamic Output Feedback Controllers}
\shortauthors{T.Sadamoto and T.Hirai}

\title [mode = title]{Policy Gradient Methods for Designing Dynamic Output Feedback Controllers}                      

\tnotetext[1]{This paper has not been presented at any conferences. Corresponding author: T. Sadamoto, Tel.: +8142-443-5172, Email: sadamoto@uec.ac.jp.}


\author[1]{Tomonori Sadamoto}[auid=000,bioid=1,
                        orcid=0000-0002-3563-3938]


\address[1]{Department of Mechanical and Intelligent Systems Engineering, Graduate School of Informatics and Engineering, The University of Electro-Communications, 1-5-1, Chofugaoka, Chofu, Tokyo 182-8585, Japan}

\author[2]{Takumi Hirai}

\address[2]{Kubota Corporation, 1-11, Takumi-cho, Sakai-shi, Osaka, 590-0908, Japan}


\fntext[fn1]{This work is partially supported by JSPS KAKENHI JP22K14276. }

\begin{abstract}
This paper proposes model-based and model-free policy gradient methods (PGMs) for designing dynamic output feedback controllers for discrete-time partially observable deterministic systems without noise. To fulfill this objective, we first show that any dynamic output feedback controller design is equivalent to a state-feedback controller design for a newly introduced system whose internal state is a finite-length input-output history (IOH). Next, based on this equivalence, we propose a model-based PGM and show its global linear convergence by proving that the Polyak-\L ojasiewicz inequality holds for a reachability-based lossless projection of the IOH dynamics. Moreover, we propose a model-free implementation of the PGM with a sample complexity analysis. Finally, the effectiveness of the model-based and model-free PGMs is investigated through numerical simulations.  
\end{abstract}




\maketitle

\section{Introduction}\label{sec_intro}
With the increasing complexity of systems to be controlled, data-driven controls \cite{camp,Jiang,de2019formulas}, especially online learning methods \cite{sutton,fazel,bu2019lqr, Mih1, Mih2,Mih3}, are gaining attention due to their memory efficiency and ease of implementation. 
Online learning methods can be divided into two categories: methods that update the value/Q functions \cite{sutton} and methods that update the policy directly \cite{fazel,bu2019lqr, Mih1, Mih2,Mih3}. Though the former can handle general control policies, the convergence speed is usually slower than that of the latter owing to its generality. The most well-known latter approach is the policy gradient method (PGM) \cite{sutton,gullapalli1990stochastic}. 
However, the convergence of the control policy to an optimal one is not guaranteed in general. Recently, \cite{fazel} has shown the global convergence of the PGM in a standard linear quadratic regulator (LQR) setting. This paper showed that the cost function in the LQR setting is non-convex but is in a class of functions satisfying the Polyak-\L ojasiewicz (PL) inequality \cite{invex}, thereby guaranteeing global convergence. Moreover, a model-free implementation of the PGM based on the Monte Carlo approximation of the gradient has also been proposed. Similar approaches have been developed in continuous-time LQR settings \cite{Mih1, Mih2,Mih3}. However, these methods assume the availability of all state variables, which prevents their application to real control systems, such as power grids.\par

From this perspective, recent works \cite{fatkhullin2021optimizing,takakura2022structured,bu2019topological,duan2023optimization2} have studied static-output-feedback (SOF) control to optimize a linear quadratic cost function. However, unlike the state-feedback case, convergence to a globally optimal solution through the PGM is not guaranteed. This is because the set of stabilizing SOF controllers is typically disconnected, and stationary points can be local minima, saddle points, or even local maxima \cite{fatkhullin2021optimizing}. To overcome this difficulty, a PGM for designing dynamic optimal output feedback controllers must be developed. 
Although one aspect of its optimization landscape has been studied for the LQR problem \cite{duan2022optimization} and LQG/$H_{\infty}$ control problems \cite{tang2021analysis,hu2022connectivity}, to the best of our knowledge, there are no concrete methodologies for designing dynamic output feedback controllers with a theoretical guarantee of global convergence.

Against this background, we propose model-free policy gradient methods for designing dynamic output feedback controllers for discrete-time, partially observable systems. As a first step, we focus on noise-free settings, though we numerically investigate the effectiveness of the proposed method in the presence of noise. The contributions of this paper are as follows. 

$\bullet$~We show that designing any dynamic output feedback controller without a feedthrough term for a given partially observable system is equivalent to designing a {\it state}-feedback gain for a new system whose internal state is a finite-length {\it input-output history (IOH)}. We refer to the new system and the gain as the {\it IOH dynamics} and {\it IOH-feedback gain}. In addition to showing the equivalence, we explain how to transform a designed IOH-feedback gain into the corresponding dynamic output feedback controller. These novel findings enable us to deal with dynamic controller designs as IOH-feedback gain designs. 

$\bullet$~We show the global linear convergence of the PGM for designing an IOH-feedback gain to an optimal gain if the gradient can be computed exactly (i.e., the IOH dynamics are known). Unlike the state-feedback case \cite{fazel}, the PL inequality of the IOH dynamics does not hold. This is because of the unavoidable singularity of the second-moment matrix of the IOH; thus, the straightforward application of the result \cite{fazel} to this design is difficult. However, in the language of control theory, this unavoidable singularity implies that the reachable subspace of IOH is limited. Based on this observation, we consider projecting the IOH dynamics onto its reachable subspace in a lossless manner and investigate the optimization landscape for these projected dynamics. As a result, we show that the PL inequality holds for the projected IOH dynamics, ensuring global linear convergence to an optimal IOH-feedback gain. 
This means that an optimal dynamic output feedback controller can be obtained through the PGM using the equivalent transformation mentioned above. To the best of our knowledge, this paper is the first to show the global linear convergence of the PGM for designing dynamic output feedback controllers. This result was achieved by considering an optimization problem over IOH-feedback gains rather than the system matrices of dynamical controllers, as in existing works such as  \cite{duan2022optimization,tang2021analysis}. Because of the difference in the parameter space for optimization problems, the insights presented in the works may not apply to our problem. For example, \cite{duan2022optimization} demonstrated that all minimal stationary points are globally optimal, whereas the dynamical controller obtained by the PGM over IOH-feedback gains may not be minimal. These results are not contradictory due to the differences in the parameter spaces. In fact, unlike in \cite{duan2022optimization}, the PL inequality holds in our problem, ensuring convergence to a globally optimal controller.

$\bullet$~We propose a model-free implementation of the PGM by using the zeroth-order approximation of the gradient \cite{fazel}. Moreover, we show a sample complexity analysis of the model-free PGM, and we present a guideline for choosing design parameters. This analysis itself is not significantly different from that in \cite{fazel}; however, we believe that it is important to show that a similar claim also holds for the dynamic output feedback controller design via the IOH framework. 

$\bullet$~The effectiveness of the proposed methods is investigated through numerical simulations. While the above theoretical investigation focuses on noise-free systems, in the numerical simulations, we investigate the effectiveness of the proposed model-free PGM in the presence of both process and observation noise. As a result, it is shown that the proposed PGM can robustly learn dynamic output feedback controllers.

{\it Organization:} Section \ref{PROBLEM SETUP} formulates the IOH-feedback controller design problem, and shows its equivalence with the dynamic output feedback controller design problem. Section \ref{proposed method} presents the PGM and its global linear convergence if the model information is hypothetically available. Section \ref{sec-model-free} presents a model-free implementation of the PGM and presents the sample complexity analysis. Section \ref{numerical simulation} shows the numerical simulation results. Finally, Section \ref{conclusion} concludes the paper. 

{\it Notation:} We denote the set of $n$-dimensional real vectors as $\mathbb{R}^n$, 
the set of natural numbers as $\mathbb N$, 
the set of positive real numbers as $\mathbb R_+$, 
the $n$-dimensional identity matrix as $I_n$, and
the $n$-by-$m$ zero matrix as $0_{n \times m}$. The subscript $n$ (resp. $n \times m$) of $I_n$ (resp. $0_{n \times m}$) is omitted if obvious. 
Given a matrix, entries with a value of zero are left blank, unless this would cause confusion. The stack of $x(t)$ for $t\in [t_1,t_2]$ is denoted as $[x]_{t_2}^{t_1} \coloneqq [x(t_1)^{\top},\cdots, x(t_2)^{\top}]^{\top}$ while the set as $\{x\}^{t_1}_{t_2}$. 
For any matrix-valued random variable $A \in \mathbb R^{n \times m}$, we denote its expectation value as ${\mathbb E}[A]$.
For any $A \in \mathbb R^{n \times m}$, the Moore--Penrose inverse as $A^{\dagger}$, minimum singular value as $\sigma_{\rm min}(A)$, 
trace as $\tr (A)$, 2-induced norm as $\|A\|$, 
Frobenius norm as $\|A\|_F$, and 
the spectral radius as $\sprad (A)$. Further, the subspace spanned by the columns of $A$ is denoted as $\im A$. 
The gradient of a differentiable function $f(\cdot): \mathbb R^{n \times m} \rightarrow \mathbb R$ at $A\in {\mathbb R}^{n\times m}$ is denoted as $\nabla f(A)$, and the Hessian of $f$ at $A$ is denoted as $\nabla^2 f(A)$. 
For any symmetric matrix $A \in{\mathbb R}^{n\times n}$, the positive (semi)definiteness of $A$ is denoted by $A>0$ ($A \geq 0$). 
For $A \geq 0$, we define $A^{\frac{1}{2}}$ such that $A = A^{\frac{1}{2}}A^{\frac{\top}{2}}$ holds. We denote the probability of satisfying a given condition $a$ as ${\mathbb P}(a)$. 
When $a \in \mathbb R^{n}$ follows a Gaussian distribution whose second moment around zero is $\Phi \in \mathbb R^{n\times n}$, we denote this fact as $a \sim \mathcal N_{\Phi}$. Note that $\mathcal N_{\Phi}$ does not specify what the mean of $a$ is; however, we will use this simplified notation when the argument does not depend on the mean. Given an $n$-dimensional system ${\bm \Sigma}: x(t+1) = Ax(t) + Bu(t)$, $y(t) = Cx(t)$ where $u \in \mathbb R^m$ and $y \in \mathbb R^r$, and $L \in \mathbb N$, we define 
$\mathcal R_L({\bm \Sigma}) \coloneqq [A^{L-1}B, \ldots, B]$,  
$\mathcal O_L({\bm \Sigma}) \coloneqq [C^{\top}, \ldots, (CA^{L-1})^{\top}]^{\top}$, and 
$\mathcal H_L({\bm \Sigma}) \coloneqq [H_{i,j}]$ where $H_{i,j} \in \mathbb R^{r \times m}$ is the $(i,j)$-th block submatrix such that $H_{i,j} = 0$ for $i \leq j$ and $H_{i,j} = CA^{i-j-1}B$ for $i > j$. 


\section{Problem Setup}\label{PROBLEM SETUP}
We consider a MIMO discrete-time system described as
\begin{equation}\label{1}
{\bm \Sigma}_{\rm s} : 
\begin{cases}
    &x(t+1) = Ax(t)+Bu(t)\\
    &y(t) = Cx(t)
\end{cases}, \quad t \geq 0,
\end{equation}
where $x\in {\mathbb R}^{n}$ is the state, $u\in {\mathbb R}^m$ is the control input, and $y\in {\mathbb R}^r$ is the evaluation output. We impose the following three assumptions on \eqref{1}.
\begin{assumption}\label{ass_ABC}
The matrices $A$, $B$, and $C$ are unknown, and ${\bm \Sigma}_{\rm s}$ is a minimal realization.
\end{assumption}
\begin{assumption}\label{ass_uy}
$x$ is not measurable, but $u$ and $y$ are.
\end{assumption}

Our objective is to design an output feedback controller 
\begin{equation}\label{dyn_K}
    {\bm K}_{\rm s} : 
\begin{cases}
    &\xi(t+1) = \Xi \xi(t) + \Lambda y(t)\\
    &u(t) = \Omega \xi(t)
\end{cases}, \quad t \geq 0,
\end{equation}
from the input-output data of ${\bm \Sigma}_{\rm s}$ to improve the sum of the quadratic cost on $y$ and $u$; the detailed definition is described in \eqref{defJ} later. Data-driven design of ${\bm K}_{\rm s}$ is more challenging than that of a state-feedback controller $u = Fx$ because its optimization landscape is more complicated, as discussed in \cite{duan2022optimization,tang2021analysis}. To overcome this difficulty, we propose a new approach that is {\it similar} to state-feedback controller designs. Some preliminaries for this purpose are introduced in the next subsection. 

\subsection{Preliminary for Formulation}\label{subsec_pre}
\begin{definition}\label{def_1}
Let $\{u,y\}$ be the input-output signal of ${\bm \Sigma}_{\rm s}$ in \eqref{1}. Given $L \in \mathbb N$, we refer to
\begin{equation}\label{def_IOH}
    v(t) \coloneqq [([u]^{t-L}_{t-1})^{\top}, ([y]^{t-L}_{t-1})^{\top}]^{\top} \in \mathbb R^{L(m+r)}, \quad t \geq L
\end{equation}
as an {\it $L$-length input-output history}, or simply, an {\it IOH}. 
\end{definition}

\begin{lem}\label{lem_VARX}
\label{IOH}
Consider ${\bm \Sigma}_{\rm s}$ in \eqref{1} and $v$ in \eqref{def_IOH}. If 
\begin{align}
\label{condition L}
    \rank{\mathcal O}_L({\bm \Sigma}_{\rm s}) = n
\end{align}
holds, then for any $u$ and $x(0)$, the IOH $v$ and output $y$ obey
\begin{equation}\label{dyn_IOH}
{\bm \Sigma} : 
\begin{cases}
    &v(t+1) = \Theta v(t) + \Pi u(t)\\
    &y(t) = C \Gamma v(t)
\end{cases}, \quad t \geq L,
\end{equation}
where $\Gamma \hspace{-0.5mm}\coloneqq\hspace{-0.5mm} \left[{\mathcal R}_L({\bm \Sigma}_{\rm s}) - A^L{\mathcal O}_L^{\dagger}({\bm \Sigma}_{\rm s}){\mathcal H}_L({\bm \Sigma}_{\rm s}), ~A^L{\mathcal O}_L^{\dagger}({\bm \Sigma}_{\rm s})\right]$, 
\begin{equation}\label{def_Theta}
    \Theta \hspace{-0.5mm}\coloneqq\hspace{-1mm} \left[\hspace{-1mm}\begin{array}{cccc}
    \hspace{-0.5mm}& \hspace{-2mm}I_{(L-1)m} & \hspace{-2mm} & \hspace{-2mm}\\
    \hspace{-0.5mm}0_{m \times m} & \hspace{-2mm}& \hspace{-2mm}& \hspace{-2mm}\\
    \hspace{-0.5mm}& \hspace{-2mm}& \hspace{-2mm} & \hspace{-2mm}I_{(L-1)r} \\ 
    \hspace{-0.5mm}& \hspace{-2mm}& \hspace{-2mm} 0_{r \times r} & \hspace{-2mm}
    \end{array}\hspace{-2mm}
\right]+ \left[\hspace{-1mm}\begin{array}{c}
0_{(L-1)m \times L(m+r)} \\ 0_{m \times L(m+r)} \\ 0_{(L-1)r \times L(m+r)} \\ C\Gamma
\end{array}\hspace{-1mm}\right]
\end{equation}
and $\Pi \coloneqq [0_{m \times (L-1)m}, I_m, 0_{m \times Lr}]^{\top}$.  Further, it follows that
\begin{equation}\label{init_v}
    v(t) \in \mathscr{P} \coloneqq \im [\Theta^{L(m+r)-1}\Pi, \ldots, \Pi], \quad t \geq L. 
\end{equation}
Moreover, ${\dim} \mathscr{P} = Lm+n$. 
\end{lem}
\begin{proof} See Appendix \ref{pf_App1}. 
\end{proof}
This lemma shows the following two facts:
\begin{enumerate}
    \item[i)] The system ${\bm \Sigma}_{\rm s}$, whose internal state (i.e., $x$) is unmeasurable, can be equivalently represented as ${\bm \Sigma}$, the internal state of which (i.e., $v$) is measurable. 
    \item[ii)] Any $v$ is included in the $(Lm+n)$-dimensional subspace $\mathscr{P}$. Note that this holds even for the initial IOH $v(L)$. This relation will be used for the problem formulation described later.  
\end{enumerate}
Owing to i), similarly to state-feedback controller designs such as the pole placement, it might be possible to improve the performance of $y$ by designing an {\it IOH-feedback} controller
\begin{equation}\label{IOH_law}
    {\bm K}:~u(t) = Kv(t), \quad t \geq L
\end{equation}
for the hypothetical system ${\bm \Sigma}$, where $K \in \mathbb R^{m \times L(m+r)}$. Note that the control law \eqref{IOH_law} begins at $t=L$, while for $t \in [0,L-1]$ the input sequence $\{u\}^{0}_{L-1}$ determined as the first to $Lm$-th elements of $v(L)$ is supposed to be already applied. Once such a control gain $K$ is found, the controller \eqref{IOH_law} can be implemented in the form of \eqref{dyn_K}, as shown below. 

\begin{lem}\label{lem2}
Given $L \in \mathbb N$ satisfying \eqref{condition L}, consider ${\bm \Sigma}$ in \eqref{dyn_IOH}. Given ${\bm K}$ in \eqref{IOH_law}, let $K$ be partitioned as $K = [A_L, \cdots, A_1$, $B_L, \cdots, B_1]$, where $A_i \in \mathbb R^{m \times m}$ and $B_i \in \mathbb R^{m \times r}$. Consider ${\bm K}_{\rm s}$ in \eqref{dyn_K} with 
\begin{equation}\label{ABCD_hat}
    \Xi \hspace{-0.5mm}=\hspace{-0.5mm} \left[\hspace{-1mm}
    \begin{array}{cccc}
    & & & A_L \\
    I  &  & & A_{L-1} \\
    & \ddots &  & \vdots \\
    & & I & A_1
    \end{array}\hspace{-1mm}
    \right]\hspace{0mm},~\Lambda \hspace{-0.5mm}=\hspace{-0.5mm} \left[\hspace{-1mm}
    \begin{array}{c}
        B_L \\ B_{L-1} \\ \vdots \\ B_1
    \end{array}\hspace{-1mm}
    \right]\hspace{0mm},~\Omega \hspace{-0.5mm}=\hspace{-0.5mm} \left[
    \begin{array}{c}
    0 \\ \vdots \\ 0 \\ I
    \end{array}
    \right]^{\top} 
\end{equation}
    and $\xi(0) = \mathcal O_L^{-1}({\bm K}_{\rm s})[I, -\mathcal H_L({\bm K}_{\rm s})] v(L)$, where $\mathcal O_L({\bm K}_{\rm s})$ is always invertible. Then, the $u$ and $y$ of the closed-loop  $({\bm \Sigma}, {\bm K})$ are identical to those of $({\bm \Sigma}_{\rm s}, {\bm K}_{\rm s})$. 
\end{lem}
\begin{proof} See Appendix \ref{pf_App2}. 
\end{proof}
This lemma shows an equivalence between the closed-loops $({\bm \Sigma}, {\bm K})$ and $({\bm \Sigma}_{\rm s}, {\bm K}_{\rm s})$. 
Based on these observations, we propose the following two-step approach for designing ${\bm K}_{\rm s}$ in \eqref{dyn_K}: 
\begin{enumerate}
    \item Design $K$ in \eqref{IOH_law} to improve the performance of $y$ in \eqref{dyn_IOH}.
    \item Transform ${\bm K}$ to ${\bm K}_{\rm s}$ in \eqref{dyn_K} by Lemma \ref{lem2}. 
\end{enumerate}
In the remainder of this paper, we focus on the step 1. Next, we show the problem formulation. 

\begin{rem}
 The dynamic controller \eqref{dyn_K} and \eqref{ABCD_hat} are always observable but not necessarily reachable. Observability is evident from the fact that $\mathcal O_L({\bm K}_{\rm s})$ is always invertible. For reachability, for example, when $n=m=r=L=1$, \eqref{ABCD_hat} becomes $\Xi = A_1$, $\Lambda = B_1$ and $\Omega = 1$. Thus, if $B_1 \not= 0$, then ${\bm K}_{\rm s}$ is reachable whereas if $B_1 = 0$ then it is unreachable. As presented in this example, the reachability of the dynamic controller depends on $K$. 
 \end{rem}

\subsection{Formulation and Overview of Approach}
\begin{prob}\label{problem_1}
Given $L \in \mathbb N$ satisfying \eqref{condition L}, consider ${\bm \Sigma}$ in \eqref{dyn_IOH}, where ${\bm \Sigma}_{\rm s}$ is supposed to satisfy Assumptions \ref{ass_ABC}-\ref{ass_uy}. Let $\Phi \geq 0$ be given such that 
\begin{equation}\label{def_psi}
    \im \Phi = \mathscr{P},
\end{equation}
where $\mathscr{P}$ is defined in \eqref{init_v}. Given $Q > 0$ and $R>0$, find $K$ in \eqref{IOH_law} such that
\begin{equation}\label{defJ}
    J(K) \coloneqq {\mathbb E}_{v(L) \sim \mathcal N_{\Phi}} \left[\sum_{t=L}^{\infty} y^{\top}(t)Qy(t)  + u^{\top}(t) Ru(t) \right]
\end{equation}
is as small as possible. 
\end{prob}
Again, note that once we obtain $K$, we can construct ${\bm K}_{\rm s}$ in \eqref{dyn_K} by Lemma \ref{lem2}. The randomization in \eqref{defJ} is needed for making the design problem initial-IOH-independent, as described in the following. 

\begin{propo}\label{prop0_1}
 Consider Problem \ref{problem_1}. Define 
 \begin{equation}\label{opt_SFB}
     K_{\rm SF}^{\star} \coloneqq -(R+B^{\top}XB)^{-1}B^{\top}XA, 
 \end{equation}
 where $X \geq 0$ is the solution of ${\rm Ric}(X) \coloneqq A^{\top}XA - X + A^{\top}XB(R + B^{\top}XB)^{-1}B^{\top}XA + C^{\top}QC = 0$. Then, the minimizer of $J$ in \eqref{defJ} has a form of $u = K^{\star}v$ where 
 \begin{equation}\label{optK}
     K^{\star} = K_{\rm SF}^{\star}(\Gamma + \bar{\Gamma}), 
 \end{equation}
 $\Gamma$ is defined in \eqref{dyn_IOH}, $\bar{\Gamma} \in \mathbb R^{n \times L(m+r)}$ satisfies $\ker \bar{\Gamma} = \mathscr{P}$, and $\mathscr{P}$ is defined in \eqref{init_v}. 
\end{propo}
\begin{proof}
 See Appendix \ref{app_propo}
\end{proof}%
This proposition implies the following two facts. First, as in the usual LQR, the minimizing gain $K^{\star}$ does not depend on the subspace $\mathscr{P}$ or the second moment $\Phi$ if \eqref{def_psi} is satisfied. 
Second, the optimal control law is equivalent to the state-feedback-type optimal control law $u=K_{\rm SF}^{\star}x$. 
If $A$, $B$, and $C$ were known, we could find $K^{\star}$ easily; however, owing to the inaccessibility of these matrices, this model-based approach is not feasible.

As an alternative, we propose a data-driven PGM, which iteratively update the gain $K$ based on a Monte Carlo approximation of the gradient of $J$. To this end, as a first step, we show a PGM assuming that the gradient is exactly computable (i.e., the model is available), and show that the method is globally linearly convergent.

\begin{rem}\label{rm_choose_L}
One way to find $L$ without using a model is to provide a relatively large value for it because no theoretical
problem occurs even if $L$ is large. Later, we will show this in numerical simulations. 
\end{rem}

\begin{rem}
As shown in the proof of Lemma \ref{lem_VARX}, the initial IOH $v(L)$ satisfies $v(L) = \mathcal P [([u]^{0}_{L-1})^{\top}, x^{\top}(0)]^{\top}$ where $\mathcal P$ is defined in \eqref{del_calP}, and $\im \mathcal P = \mathscr{P}$. Therefore, as long as the second moment of $[([u]^{0}_{L-1})^{\top}, x^{\top}(0)]^{\top}$ is positive definite, for example, if $x(0)$ and $u(0), \ldots, u(L-1)$ are i.i.d. Gaussian samples, the condition \eqref{def_psi} can be satisfied. Even though $x(0)$ cannot be measured, the situation where $x(0)$ follows some distribution can occur such as by setting the target system to a certain initial situation. Therefore, the assumption \eqref{def_psi} would be practical.    
\end{rem}

\section{Model-Based Policy Gradient Method and its Convergence Analysis}\label{proposed method}

\subsection{Model-Based Policy Gradient Method}\label{MB_PGM}
The PGM is described as
\begin{equation}
\label{gd}
    K_{i+1} = K_{i}-\alpha\nabla J(K_i),
\end{equation}
where $i \geq 0$ is an iteration number and $\alpha \in \mathbb R_+$ is a given step-size parameter. We derive an explicit representation of $\nabla J(K_i)$. To fulfill this objective, we characterize the stability of the closed-loop of ${\bm \Sigma}$ in \eqref{dyn_IOH} with ${\bm K}_i:u(t) = K_i v(t)$ for $t \geq L$, described as
\begin{equation}\label{defTHK}
    ({\bm \Sigma}, {\bm K}_i):~v(t+1) = \Theta_{K_i} v(t), \quad \Theta_{K_i} \coloneqq \Theta + \Pi K_i
\end{equation}
for $t\geq L$. Note that the stability of $\Theta_{K_i}$ is sufficient but not necessary for $v(\infty) = 0$ because $v \in \mathbb R^{L(m+r)}$ is constrained on the $(Lm+n)$-dimensional subspace $\mathscr{P}$ in \eqref{init_v}. In other words, the stability of $v$ over $\mathscr{P}$ is necessary and sufficient for $v(\infty) = 0$. To see this more clearly, we introduce the following lemmas.
\begin{lem}\label{lem_DYC}
 Consider ${\bm \Sigma}$ in \eqref{dyn_IOH} and $\mathscr{P}$ in \eqref{init_v}. Then, there exists $P \in \mathbb R^{L(m+r) \times (Lm+n)}$ such that 
\begin{equation}\label{prop_P}
    \im P = \mathscr{P}, \quad P^{\top}P = I. 
\end{equation}
Further, $v$ follows 
\begin{equation}\label{dyn_IOH2}
\begin{cases}
    &\hat{v}(t+1) = P^{\top}\Theta P \hat{v}(t) + P^{\top}\Pi u(t)\\
    &v(t) = P \hat{v}(t)
\end{cases}, \quad t \geq L,
\end{equation}
where $\hat{v}(L) = P^{\top}v(L)$. Besides, if $v(L) \sim \mathcal N_{\Phi}$, where $\Phi$ satisfies \eqref{def_psi}, then 
\begin{equation}\label{def_Phihat}
    \hat{v}(L) \sim \mathcal N_{\hat{\Phi}}, \quad \hat{\Phi} \coloneqq P^{\top}\Phi P > 0. 
\end{equation}
\end{lem}
\begin{proof} 
See Appendix \ref{pf_App3}. 
\end{proof}
\begin{lem}\label{PTP_lem}
Consider Problem \ref{problem_1}, $({\bm \Sigma}, {\bm K}_i)$ in \eqref{defTHK}, and $P$ in \eqref{prop_P}. Then $v(\infty) = 0$ for any $v(L) \sim \mathcal N_{\Phi}$ if and only if $P^{\top}\Theta_{K_i}P$ is Schur. 
\end{lem}
\begin{proof}
The claim immediately follows from \eqref{dyn_IOH2} and the fact that $u(t) = K_i v(t) = K_iP\hat{v}(t)$ for $t \geq L$.
\end{proof}


Based on these lemmas, we derive an explicit representation of $\nabla J$ in \eqref{gd}. Define
\begin{equation}\label{def_THhat}
    \hat{K}_i \coloneqq K_iP, \quad \hat{\Gamma} \coloneqq \Gamma P, \quad \hat{\Theta}_{K_i} \coloneqq P^{\top}(\Theta + \Pi K_i)P,
\end{equation}
where $\Gamma$ is defined in \eqref{dyn_IOH}. If $\hat{\Theta}_{K_i}$ is Schur, it follows from Lemma~\ref{PTP_lem} that there exists $\hat{\Psi}_{K_i} \geq 0$ satisfying
\begin{equation}\label{5}
   \hat{\Theta}_{K_i}^{\top}\hat{\Psi}_{K_i}\hat{\Theta}_{K_i} -\hat{\Psi}_{K_i} = -\hat{\Gamma}^{\top}C^{\top}QC\hat{\Gamma}-\hat{K}_i^{\top}R\hat{K}_i,
\end{equation}
where $Q$ and $R$ are defined in \eqref{defJ}. Then, the cost $J$ in \eqref{defJ} can be written as 
\begin{equation}\label{6}
J(K_i) = {\mathbb E}\left[\hat{v}^{\top}(L)\hat{\Psi}_{K_i}\hat{v}(L)\right] = {\mathbb E}\left[v^{\top}(L)\Psi_{K_i}v(L)\right],
\end{equation}
where 
\begin{equation}\label{def_PSI}
    \Psi_{K_i} \coloneqq P\hat{\Psi}_{K_i}P^{\top}.
\end{equation}
Under these settings, $\nabla J$ is given by the following lemma.
\begin{lem}\label{gradient}
Consider the closed-loop $({\bm \Sigma}, {\bm K}_i)$ in \eqref{defTHK}. Assume $\hat{\Theta}_{K_i}$ in \eqref{def_THhat} is Schur. Define 
\begin{equation}\label{def_EV}
\begin{cases}
    &E_{K_i} \hspace{-0.5mm}\coloneqq\hspace{-0.5mm} (R\hspace{-0.5mm}+\hspace{-0.5mm}\Pi^{\top}\Psi_{K_i}\Pi)K_i\hspace{-0.5mm}+\hspace{-0.5mm}\Pi^{\top}\Psi_{K_i}\Theta,\\
    & V_{K_i} \hspace{-0.5mm}\coloneqq\hspace{-0.5mm} {\mathbb E}\left[\sum_{t = L}^{\infty}v(t)v^{\top}(t)\right]\hspace{-1mm}
\end{cases},
\end{equation}
where $v$ follows \eqref{defTHK}. Then $\nabla J$ in \eqref{gd} is described as 
\begin{equation}\label{8}
    \nabla J(K_i) = 2E_{K_i}V_{K_i}. 
\end{equation}
\end{lem}
\begin{proof}
See Appendix \ref{app_lemcc}. 
\end{proof}%

\subsection{Convergence Analysis}\label{CONVERGENCE ANALYSIS}
\begin{lem}\label{lem_stab}
 Consider \eqref{dyn_IOH2}, $J$ in \eqref{defJ}, and $\hat{\Theta}_{K}$ in \eqref{def_THhat}. Then, $\hat{\Theta}_{K}$ is Schur if and only if $J(K)$ is bounded.
\end{lem}
\begin{proof}
See Appendix~\ref{app_lem_stab}. 
\end{proof}
For the following analysis, we define a sublevel set of stabilizing controllers as 
\begin{equation}\label{def_Kdom}
    \mathbb K \coloneqq \{K~|~ J(K) \leq c,~J~{\rm is~defined~as}~\eqref{defJ}\},
\end{equation}
where $c$ is a given sufficiently large scalar satisfying $c \geq J(K_0)$ with given $K_0$ such that $\hat{\Theta}_{K_0}$ in \eqref{def_THhat} is Schur. 
The cost function $J$ and the set $\mathbb K$ may not be convex, which is in fact true even for the state-feedback case \cite{fazel}. Instead, following \cite{fazel}, to show the convergence of the gradient algorithm \eqref{gd} with \eqref{8}, we will use the fact that a gradient algorithm with an appropriately chosen step-size parameter is globally linearly convergent if the objective function is smooth and satisfies the {\it PL inequality} \cite{PLinequality}. The PL inequality of $J$ over $\mathbb K$ is shown by the following lemma.

\begin{lem}\label{PL Lem MIMO}
Consider Problem \ref{problem_1}, $K^{\star}$ in \eqref{optK}, and $\mathbb K$ in \eqref{def_Kdom}. Let $\hat{v}$ follow \eqref{dyn_IOH2}, where $u$ is given as \eqref{IOH_law}. Then the PL inequality 
\begin{equation}\label{PL MIMO}
    J(K) - J(K^{\star}) \leq \frac{\|\hat{V}_{K^{\star}}\|}{4\sigma_{\rm min}(R)\sigma_{\rm min}^2({\hat{V}_{K}})}\|\nabla J(K)\|_F^2,
\end{equation}
holds for any $K \in \mathbb K$ where $\hat{V}_K \coloneqq {\mathbb E}\left[\sum_{t = L}^{\infty} \hat{v}(t)\hat{v}^{\top}(t)\right] > 0$. 
\end{lem}

\begin{proof}
See Appendix \ref{SISO Case}. 
\end{proof}

It should be emphasized that the reachability-based reduced-order representation \eqref{dyn_IOH2} plays an important role for deriving \eqref{PL MIMO}. Because $\hat{v}$ is the projection of $v$ onto its reachable subspace $\mathscr{P}$ in \eqref{init_v}, the matrix $\hat{V}_K$ is invertible. Hence, the term $\sigma_{\rm min}(\hat{V}_K)$ becomes nonzero; as a result, \eqref{PL MIMO} holds.
If we were to apply the argument in \cite{fazel} straightforwardly to the IOH dynamics \eqref{dyn_IOH}, this term would be $\sigma_{\rm min}(V_K)$ with $V$ defined in \eqref{def_EV}, which must be zero due to \eqref{init_v}. To avoid this singularity, we used the reachability-based representation \eqref{dyn_IOH2}, thereby successfully showing the PL inequality for $J$. The smoothness of $J$ over the sublevel set $\mathbb K$ is shown in the following lemma.  

%

\begin{lem}\label{smoothness_new}
Consider Problem \ref{problem_1} and $\mathbb K$ in \eqref{def_Kdom}.
Given $K\in {\mathbb K}$, it follows that
\begin{equation}\label{def_qqq}
J(K')- J(K) \leq \tr\left(\nabla J(K)(K'-K)^{\top}\right) + \frac{q}{2}\|K'-K\|_F^2
\end{equation}
for any $K' \in \mathbb K$, where 
\begin{eqnarray}
&& \hspace{-15mm} q  \coloneqq 2\left(\frac{c}{\sigma_{\rm min}(\hat{\Phi})} +\|R\|\right)\frac{Lc}{\rho} + 4 \frac{\sqrt{2(Lm +n)}}{\sigma_{\rm min}(\hat{\Phi})} \nonumber \\
        &&  \hspace{-12mm} \times \left(\frac{2c}{\sigma_{\rm min}(\hat{\Phi})}+\tr(R) -\tr(\hat{Q})\right)\left(\frac{Lc}{\rho}\right)^{2}, \label{q} \\ 
&& \hspace{-15mm} \scalebox{0.92}{$\displaystyle \rho \coloneqq {\rm min}(\sigma_{\rm min}(Q), \sigma_{\rm min}(R)),~ \hat{Q} \coloneqq P^{\top}\Gamma^{\top}C^{\top}QC\Gamma P $} \label{def_rhoQ}
\end{eqnarray}
and $\hat{\Phi}$ is defined in \eqref{def_Phihat}. 
\end{lem}
\begin{proof}
See Appendix \ref{app_smoothness_new}. 
\end{proof}

Based on Lemmas \ref{PL Lem MIMO}-\ref{smoothness_new} and \cite{PLinequality}, the following holds. 

\begin{theo}\label{linear_convergence}
Consider Problem \ref{problem_1}, $K^{\star}$ in \eqref{optK}, the algorithm \eqref{gd}, and $\mathbb K$ in \eqref{def_Kdom}, where $K_0$ is given such that $\hat{\Theta}_{K_0}$ in \eqref{5} is Schur. If $\alpha$ in \eqref{gd} is chosen such that $\alpha \in (0,2/q)$, then $K_{i+1}\in \mathbb K$. Moreover, it follows that
\begin{equation}\label{lin_conv}
    J(K_{i+1}) - J(K^{\star})
    \leq \beta(K_i) (J(K_i)-J(K^{\star})),
\end{equation}
where 
\begin{equation}\label{gamma_p}
\beta(K_i) \coloneqq 1-\frac{4\sigma_{\rm min}(R)\sigma_{\rm min}^2(\hat{V}_{K_i})}{\|\hat{V}_{K^{\star}}\|}\left(\alpha-\frac{q}{2}\alpha^2\right) < 1
\end{equation}
with $q$ in \eqref{q} and $\hat{V}_K$ in \eqref{PL MIMO}. 
\end{theo}
\begin{proof}
Since ${\mathbb K}$ is compact, similarly to the discussion below Lemma 3 in \cite{Mih3}, $K_{i+1} \in \mathbb K$ follows for any $\alpha \in (0, 2/q)$. Hence, \eqref{def_qqq} holds for $K \leftarrow K_i$ and $K' \leftarrow K_{i+1}$. From Theorem 4 of \cite{PLinequality} and \eqref{PL MIMO}, we have \eqref{lin_conv}. This completes the proof.
\end{proof}

In conclusion, the PGM \eqref{gd} with \eqref{8} is globally linearly convergent if $\alpha$ is sufficiently small such that $\alpha \in (0,2/q)$. Once it is converged, we can construct a dynamical controller ${\bm K}_{\rm s}$ in \eqref{dyn_K} based on Lemma~\ref{lem2}.

\begin{rem}
Dynamic output-feedback controller design using the PGM has been studied recently in literature such as \cite{duan2022optimization,tang2021analysis}. However, these studies do not propose specific algorithms ensuring global convergence because the optimization problem, which explores the system matrices of the dynamic controller (i.e., $\Xi$, $\Lambda$, and $\Omega$ in \eqref{dyn_K}), has multiple local optima as shown in those studies. By contrast, Theorem \ref{linear_convergence} shows that the PGM over IOH gains converges to a globally optimal solution because the PL inequality holds for the optimization problem whose decision parameter is taken as the IOH-gain, rather than the system matrices. In other words, through the new parameterization of dynamic controllers (i.e., IOH-gain representation), we guaranteed global convergence. This is the technical differences between our approach and existing studies.
\end{rem}


\section{Model-Free Policy Gradient Method and Sample Complexity Analysis}\label{sec-model-free}

\subsection{Model-Free PGM}\label{sec_MF_A}
We propose a model-free PGM described as 
\begin{equation}\label{modelfree}
    \tilde{K}_{i+1} = \tilde{K_{i}}- \alpha \tilde{\nabla} J(\tilde{K_i}),
\end{equation}
where $i$ is an iteration number, $\alpha$ is the step-size parameter, $\tilde{K}_i$ is the updated gain, and $\tilde{\nabla}J(\cdot) \in \mathbb R^{m \times L(m+r)}$ is an approximant of the gradient. Note that $\tilde{\nabla}J$ is a matrix to be computed from data, but is not the one by operating $\tilde{\nabla}$ to $J$. Though \eqref{modelfree} has the same form of the exact case \eqref{gd}, we use \eqref{modelfree} with the new symbols $\tilde{K}$ and $\tilde{\nabla}J$ for clarifying that those are approximants of $K$ and $\nabla J$ in \eqref{gd}. 

Let $s$ be the number of {\it episodes}, each of which refers to a finite-length sequence of the IOH starting from $t=L$, while $N$ be the length of each episode. For any $i$ in \eqref{modelfree}, $j \in \{0, \ldots, s-1\}$, and given $\tilde{K}_i$, we consider an exploration gain defined as   
\begin{align}\label{sample policy}
\tilde{K}_{i,j} \coloneqq \tilde{K}_i + \delta U_j, \quad j \in \{0,\ldots, s-1\},
\end{align}
where $U_j \in \mathbb R^{m \times L(m+r)}$ is randomly chosen such that each of its elements follows a uniform distribution while $\|U_j\|_F = 1$, and $\delta \in \mathbb R_+$ is a given constant representing the magnitude of the exploration. In addition, we denote the IOH and the increment of $J$ when the gain $\tilde{K}_{i,j}$ is used by 
\begin{eqnarray}
 \hspace{-3mm} v_{i,j}(t) &\hspace{-2mm} \coloneqq&\hspace{-2mm}
     [([u_{i,j}]^{t-L}_{t-1})^{\top}, ([y_{i,j}]^{t-L}_{t-1})^{\top}]^{\top},\\
 \hspace{-3mm} c_{i,j}(t) &\hspace{-2mm} \coloneqq&\hspace{-2mm}y_{i,j}^{\top}(t)Q y_{i,j}(t)  + u_{i,j}^{\top}(t)Ru_{i,j}(t) \label{reward_multiple}
\end{eqnarray}
for $t \geq L$ respectively, where $y_{i,j}(t)$ is the output of ${\bf \Sigma}$ in \eqref{dyn_IOH} with $u_{i,j}(t) \coloneqq \tilde{K}_{i,j}v_{i,j}(t)$. Assuming that $\{u_{i,j}, y_{i,j}\}^{L}_{L+N-1}$ is given, a zeroth-order approximant $\tilde{\nabla}J$ can be given as 
\begin{equation}\label{estimated gradient}
    \tilde{\nabla} J(\tilde{K}_i) \coloneqq \frac{mL(m+r)}{s\delta}\sum_{j=0}^{s-1}\sum_{t=L}^{L+N-1}c_{i,j}(t) U_{j}. 
\end{equation}
The pseudo-code is summarized in {\bf Algorithm 1}.  

\begin{algorithm}[h]
    \caption{\hspace{-1mm}{\bf : Model-free PGM}}
    \label{alg1}
    {\bf Initialization:}~ 
      Let $i=0$. Give $L$ satisfying \eqref{condition L}, $\tilde{K}_0$ such that $\hat{\Theta}_{\tilde{K}_0}$ in \eqref{5} is Schur, $s \in \mathbb N$, $N \in \mathbb N$, and sufficiently small $\delta, \alpha \in \mathbb R_+$.  \\
    {\bf Gradient Estimation:}\\
    {\bf for} $j=0, \ldots, s-1$, {\bf do} \vspace{-2mm}
      \begin{itemize}
          \item[1)] Give $\tilde{K}_{i,j}$ in \eqref{sample policy}.  \vspace{-2mm}
          \item[2)] Collect $\{y_{i,j}\}^{L}_{L+N-1}$ following ${\bf \Sigma}$ in \eqref{dyn_IOH} under $u(t) = u_{i,j}(t) = \tilde{K}_{i,j}v_{i,j}(t)$. \vspace{-2mm} 
          \item[3)] Compute $c_{i,j}(t)$ in \eqref{reward_multiple} for $t \in \{L, \ldots, L+N-1\}$. \vspace{-2mm}
          \item[4)] Apply $u$ such that $v(t) \sim \mathcal N_{\Phi}$. Let $t \leftarrow L$. \vspace{-2mm}
      \end{itemize}
      {\bf end}\\
      {\bf Policy Update:} \vspace{-2mm}
      \begin{itemize}
          \item[5)] Compute $\tilde{\nabla} J$ in \eqref{estimated gradient}. Next, construct $\tilde{K}_{i+1}$ by \eqref{modelfree}. \vspace{-2mm}
          \item[6)] Let $i \leftarrow i+1$ and $\tilde{K}_i\leftarrow\tilde{K}_{i+1}$. Exit if converged; otherwise, return to {\bf Gradient Estimation}. \vspace{-2mm}
      \end{itemize}
 {\bf Closing Procedure:} Let $K \leftarrow \tilde{K}_i$. Return ${\bm K}_{\rm s}$ in \eqref{dyn_K} with \eqref{ABCD_hat}. 
\end{algorithm}

\begin{rem}
As the designed IOH gain $\tilde{K}_i$ and its corresponding dynamic controller ${\bm K}_{\rm s}$ are equivalent, as shown in Lemma \ref{lem2}, one may consider that the conversion in the Closing Procedure is not necessary. The primary reason for the conversion is to clearly state that dynamic controllers in the form of \eqref{dyn_K} can be designed by the PGM through the IOH framework. Another minor reason is that once converted to a dynamic controller, we can easily use any tools developed for linear systems. For example, by applying model reduction techniques such as balanced truncation \cite{antoulas2005approximation} to the dynamic controller, we can obtain a lower-dimensional controller whose performance is almost the same as the original one if the reduction error is negligibly small. The resultant controller is easy to implement. 
\end{rem}


\subsection{Sample Complexity Analysis}\label{sec_MF_B}
We first show a sufficient condition for ensuring the stability by the exploration gain $\tilde{K}_{i,j}$ in \eqref{sample policy} and show a stochastic upper bound of the difference between the approximated gradient $\tilde{\nabla}J$ in \eqref{estimated gradient} and the exact gradient $\nabla J$ in \eqref{8}. 



\begin{lem}\label{lem_stochastic analysis}
   Consider Problem 1, ${\mathbb K}$ in \eqref{def_Kdom}, and the algorithm \eqref{modelfree} with \eqref{estimated gradient}. Let $v_{\rm max} \in \mathbb R$ be given such that $\|v(L)\| \leq v_{\max}$ almost certainly, where $v(L) \sim \mathcal N_{\Phi}$. 
If $\tilde{K}_i \in{\mathbb K}$, then the following claims hold: 

{\bf i)}~$\hat{\Theta}_{\tilde{K}_{i,j}}$ is Schur for any $j\in [0,s-1]$ and $\delta \in (0, \delta_{\rm st}]$, where 
   \begin{eqnarray}
       \label{delta_st}
       \hspace{-5mm}\delta_{\rm st} &\hspace{-3mm} \coloneqq&\hspace{-3mm} \scalebox{0.95}{$\displaystyle \min \left(
          \frac{J(\tilde{K}_i)}{\epsilon(\tilde{K}_i)},
          \frac{c-J(\tilde{K}_i)}{\epsilon(\tilde{K}_i)},
\frac{\rho\sigma_{\min}(\hat{\Phi})}{4Lc(Lm +n)h_{\tilde{K}_i}}, 2\|\tilde{K}_i\|
          \right) $} \nonumber \\
          && \\
        \epsilon(K) &\hspace{-3mm} \coloneqq&\hspace{-3mm}
 \frac{4Lc(Lm +n)\tr(\hat{\Phi})\|K\|\|R\|}{\rho\sigma_{\min}(\hat{\Phi})}\nonumber \\
    &&\hspace{-15mm}+ \frac{4(Lc(Lm +n))^2\tr(\hat{\Phi})h_K\|(\|\hat{Q}\| +9\|K\|^2\|R\|)}{\rho^2\sigma^2_{\min}(\hat{\Phi})}, \label{epsilon}  
   \end{eqnarray}
   $h_K \coloneqq \|\hat{\Theta}_K\| + \|K\|$, $\rho$ and $\hat{Q}$ are in \eqref{def_rhoQ}, $\hat{\Theta}_K$ is in \eqref{def_THhat}, and $\hat{\Phi}$ is in \eqref{def_Phihat}. 

{\bf ii)} For any $o>0$, it follows that 
   \begin{equation}
       \label{stoc_eq}
       \begin{split}
          {\mathbb P}\left(
           \|\tilde{\nabla}J(\tilde{
           K}_i) - \nabla J(\tilde{K}_i)\|_F
           \leq \theta_o
           \right)
           \geq {\rm Pr}(\tilde{K}_i, o)
       \end{split},
   \end{equation}
   where 
   \begin{eqnarray}
   \hspace{-9mm} \theta_o &\hspace{-2.5mm}\coloneqq&\hspace{-2.5mm} o +q\delta + \chi(\tilde{K}_i) \label{def_thetao}\\
   \hspace{-9mm} \chi(\tilde{K}_i) &\hspace{-2.5mm} \coloneqq&\hspace{-2.5mm}
   \textstyle \frac{m(m+r)^2L^4c^2(\|\hat{Q}\|+ 9\|\tilde{K}_{i}\|^2\|R\|)v_{\max}^2}{N\delta\sigma^2_{\min}(\hat{\Phi})\rho^2} \label{chi} \\
       \hspace{-9mm}{\rm Pr}(\tilde{K}_i,o) &\hspace{-2.5mm} \coloneqq&\hspace{-2.5mm} \scalebox{0.90}{$\displaystyle  \textstyle 1-\left(m +L(m+r)\right){\rm exp}\left(
    \frac{-so^2/2}{\zeta^2(\tilde{K}_i) + 2\zeta(\tilde{
    K}_i)o/3} \right) $} \label{Prob} \\
     \hspace{-9mm}\zeta(\tilde{K}_i) &\hspace{-2.5mm} \coloneqq&\hspace{-2.5mm} 
     \textstyle \frac{2m^{3/2}L(m+r) v_{\max}^2 J(\tilde{K}_{i})}{\delta\sigma_{\min}(\hat{\Phi})},     \label{zeta} 
   \end{eqnarray}
   with $q$ in \eqref{q}.
   \end{lem}
 
\begin{proof}
See Appendix \ref{stochastic analysis}. 
\end{proof}

Claim i) in this lemma allows us to obtain a bounded value of $c_{i,j}$ in \eqref{reward_multiple}, which is necessary for the random search. In \eqref{stoc_eq}, 
even if the error $\|\tilde{\nabla}J - \nabla J\|_F$ is large, by taking a smaller step-size $\alpha$ in \eqref{modelfree}, the two gains updated by $\tilde{\nabla}J$ and $\nabla J$ are closer to each other, i.e., $\tilde{K}_{i+1} \approx \check{K}^{[\alpha]}$, where
\begin{equation}\label{defK'}
     \check{K}^{[\alpha]} \coloneqq \tilde{K}_i -\alpha \nabla J(\tilde{K}_i).
\end{equation}
Furthermore, Theorem \ref{linear_convergence} implies that $\check{K}^{[\alpha]}$ when $\alpha$ is sufficiently small can make $J$ smaller. Therefore, in conclusion, $\tilde{K}_{i+1}$ is also expected to decrease $J$. This is, in fact, true, as shown in the following theorem.

\begin{theo}\label{linear convergent some trajectory}
Consider Problem \ref{problem_1}, $K^{\star}$ in \eqref{optK}, ${\mathbb K}$ in \eqref{def_Kdom}, $\check{K}^{[\alpha]}$ in \eqref{defK'}, and the algorithm \eqref{modelfree} with \eqref{estimated gradient}. Assume $\tilde{K}_i \in {\mathbb K}$ and $\delta$ in \eqref{sample policy} satisfy $\delta \in (0, \delta_{\rm st}]$ being with $\delta_{\rm st}$ in \eqref{delta_st}. Consider
\begin{eqnarray}
\hspace{1mm}\epsilon^*(K) &\hspace{-3mm}\coloneqq&\hspace{-3mm} \sigma_{\min}(\hat{\Phi})\|E_{K}\|^2_F \|R + \Pi^{\top}\Psi_{K}\Pi\|^{-1} \nonumber \\
\hspace{-8mm}g_{\alpha}(o) &\hspace{-3mm}\coloneqq&\hspace{-3mm} \textstyle \min \left(\hspace{-0.5mm}
    \frac{2}{q}, \frac{(1-\beta(\tilde{K}_{i}))\epsilon^*(\tilde{K}_i)}{2\epsilon(\check{K}^{[\alpha]})\theta_o}, 
    \frac{c-J(\check{K}^{[\alpha]})}{\epsilon(\check{K}^{[\alpha]})\theta_o}, \right. \nonumber \\
    && \left. \hspace{20mm} \textstyle \frac{\rho\sigma_{\min}(\hat{\Phi})}{4Lc\theta_o(Lm +n)h_{\check{K}^{[\alpha]}}}, 
    \frac{2\|\check{K}^{[\alpha]}\|}{\theta_o} 
    \hspace{-0.5mm}\right) \nonumber \\
\hspace{-8mm}\bar{o}(\alpha) &\hspace{-3mm}\coloneqq&\hspace{-3mm} \max_{o \in \mathbb R_+} o~~{\rm s.t.}~\alpha < g_{\alpha}(o) \label{def_obar},
\end{eqnarray}
where $E_{K}$ is defined in \eqref{def_EV}, $\Psi_K$ in \eqref{def_PSI}, $q$ in \eqref{q}, $\beta$ in \eqref{gamma_p}, $\rho$ in \eqref{def_rhoQ}, $\hat{\Phi}$ in \eqref{def_Phihat}, $h_K$ in Lemma~\ref{lem_stochastic analysis}, and $\theta_o$ in \eqref{def_thetao}. Then, the following two claims hold:

{\bf i)} $\lim_{\alpha \rightarrow +0} \bar{o}(\alpha) = \infty$. Further, $\bar{o}$ is continuous around a positive neighborhood at $\alpha = 0$. 

{\bf ii)} For any $s, N \in \mathbb N$, and $\alpha$ such that $\bar{o}(\alpha)$ exists, we have
\begin{align}
    &{\mathbb P}\left(
    J(\tilde{K}_{i+1}) - J(K^{\star}) \leq 
    \tilde{\beta}(\tilde{K}_i)((J(\tilde{K}_i)-J(K^{\star})) \right) \nonumber \\
    &\hspace{45mm} \geq {\rm Pr}(\tilde{K}_i, \bar{o}(\alpha)) \label{theo2_eq_1}
\end{align}
with $\tilde{\beta}(\tilde{K}_i) \coloneqq (1+\beta(\tilde{K}_i))/2 < 1$, 
where ${\rm Pr}$ is defined in \eqref{Prob}. 
\end{theo}

\begin{proof}
See Appendix \ref{linear conv modelfree}. 
\end{proof}

Claim i) shows that $\bar{o}(\alpha)$ becomes larger as $\alpha$ becomes smaller. This property is not only a sufficient condition for ensuring Claim ii), but also guarantees the improvement of ${\rm Pr}$ in \eqref{theo2_eq_1}. Claim ii) shows that the cost can be decreased by the controller updated by the model-free algorithm \eqref{modelfree} and \eqref{estimated gradient} with a probability of at least ${\rm Pr}$. Note that the stability by using the updated gain $\tilde{K}_{i+1}$ is also guaranteed with the probability at least ${\rm Pr}$; see \eqref{thm3_stab} in the proof.
As insights obtained from this theorem, we summarize the guidelines for selecting design parameters (i.e., $\alpha, s, N, \delta$) on how they impact the learning process (i.e., the improvement probability ${\rm Pr}$ and the improvement rate $\tilde{\beta}$) below. 

$\bullet$ {\it Dependency on $\alpha$:}~The step-size $\alpha$ is the most critical design parameter. This is because, as $\alpha$ decreases, $\bar{o}(\alpha)$ increases, as a result of which, the improvement probability increases (i.e., ${\rm Pr} \rightarrow 1$ as $\alpha \rightarrow +0$); however, the rate of improvement becomes slower (i.e., $\tilde{\beta} \rightarrow 1$ as $\alpha \rightarrow +0$). 


$\bullet$ {\it Dependency on $s$:}~As $s$ increases, ${\rm Pr}$ in \eqref{Prob} monotonically increases. This is intuitively evident from the fact that $s$ represents the number of Monte Carlo averages, but it can also be mathematically demonstrated by \eqref{Prob}. 

$\bullet$ {\it Dependency on $N$:}~Choosing larger value of $N$ makes ${\rm Pr}$ larger; however, the influence of this choice on ${\rm Pr}$ is limited because $\lim_{N \rightarrow \infty} \theta_o(N) = o + q\delta$, where $\theta_o(N)$ is defined in \eqref{def_thetao}. The increase of ${\rm Pr}$ is verified as follows: Denote $g_{\alpha}$ as $g_{\alpha}(o; N)$ for clarifying the dependency of $g_{\alpha}$ on $N$ through $\chi$ in \eqref{chi}. Clearly, if $N_1 \leq N_2$ then $g_{\alpha}(o; N_1) \leq g_{\alpha}(o; N_2)$. Hence, $\bar{o}(\alpha; N_1) \leq \bar{o}(\alpha; N_2)$, implying the increase of ${\rm Pr}$ in \eqref{Prob}. 

$\bullet$ {\it Dependency on $\delta$:}~As $\delta$ increases, we can see from \eqref{Prob}-\eqref{zeta} that ${\rm Pr}$ increases. However, $\delta$ should be less than $\delta_{\rm st}$ in \eqref{delta_st}; thus, the impact of choosing $\delta$ on improving the probability is limited. 

$\bullet$ {\it Dependency on the stability of closed-loop:}~As the iterations go on, $J(\tilde{K}_i)$ decreases with high probability. This yields a decrease of $\zeta$ in \eqref{zeta}, as a result of which ${\rm Pr}$ increases.  



\vspace{0mm}
\section{Numerical Simulation}\label{numerical simulation}
\begin{figure}[t]  \begin{center}
    \includegraphics[width=8.5cm]{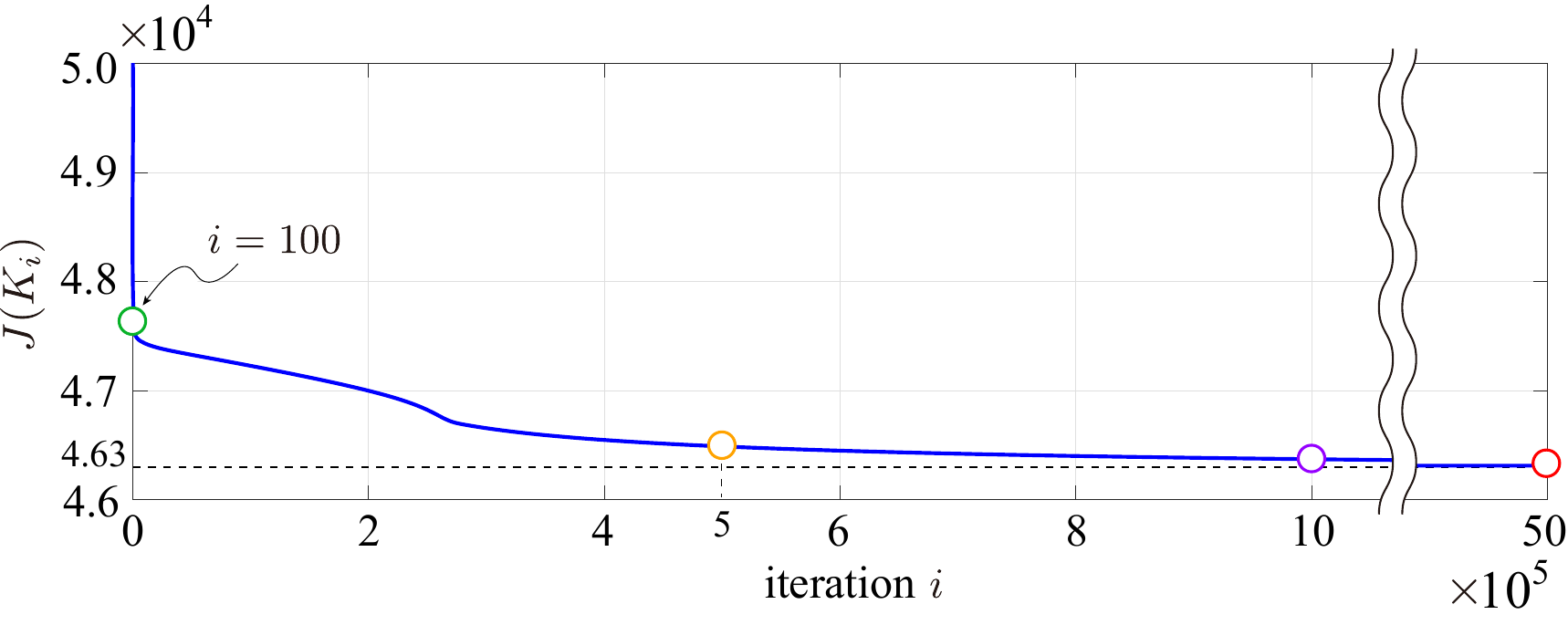} \vspace{-2mm}
    \caption{(Blue solid line) Variation of $J(K_i)$ in \eqref{defJ} for the iteration $i$ of \eqref{gd} when $L=2$. (Black dotted line) $J(K^{\star})$, where $K^{\star}$ is given by \eqref{optK}. } 
    \label{fig_J_MB} 
  \end{center}\vspace{-4mm}
\end{figure}
\begin{figure}[t]
  \begin{center}
    \includegraphics[width=8.5cm]{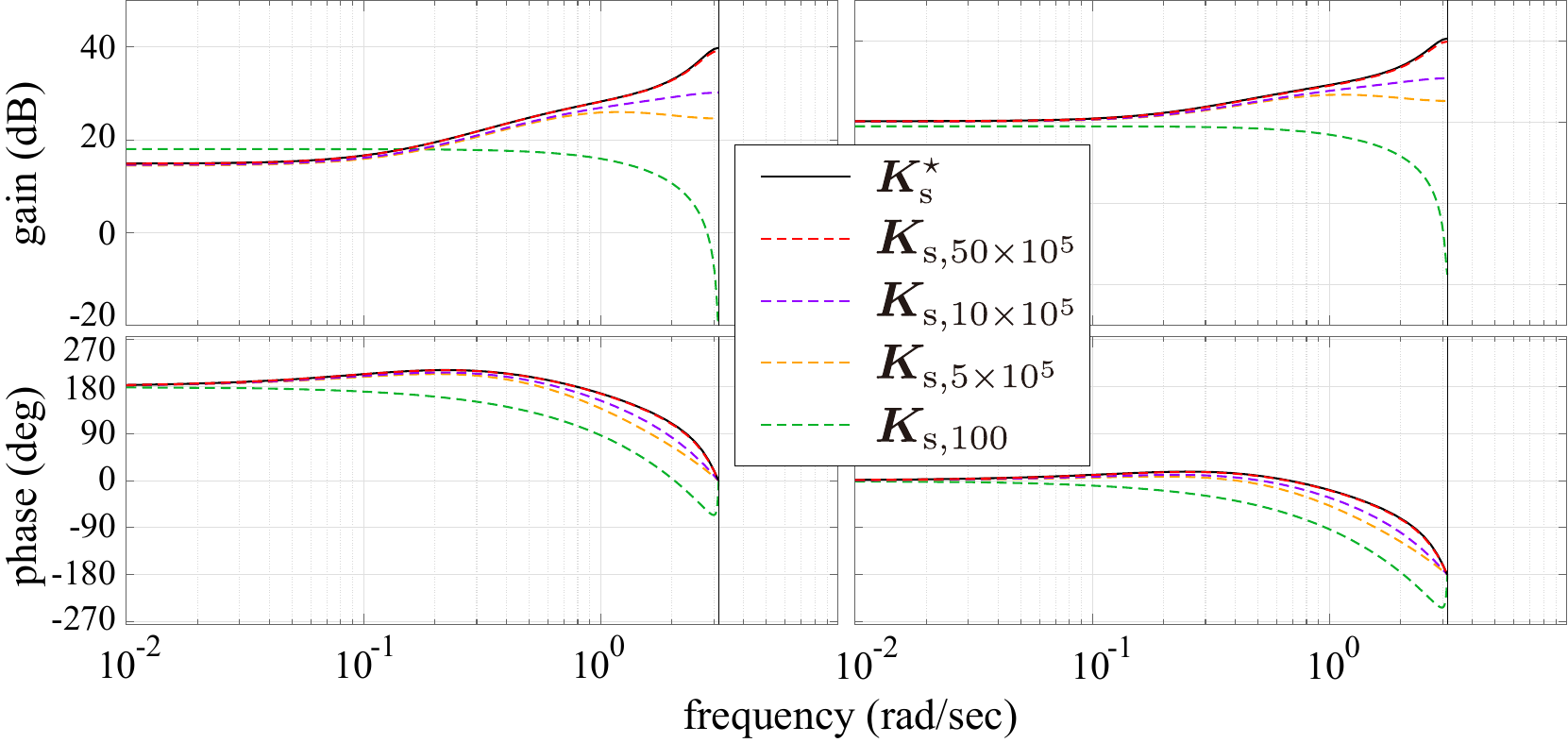} \vspace{-2mm}
    \caption{Bode diagrams of ${\bm K}^{\star}_{\rm s}$ and ${\bm K}_{{\rm s},i}$ for $i$ indicated by the circles in Figure~\ref{fig_J_MB}, where ${\bm K}^{\star}_{\rm s}$ and ${\bm K}_{{\rm s},i}$ are defined as in \eqref{dyn_K} with \eqref{ABCD_hat} for $K^{\star}$ and the corresponding $K_i$, respectively. } 
    \label{fig_Ks_MB} 
  \end{center}\vspace{-4mm}
\end{figure}
\begin{figure}[t]
  \begin{center}
    \includegraphics[width=8.5cm]{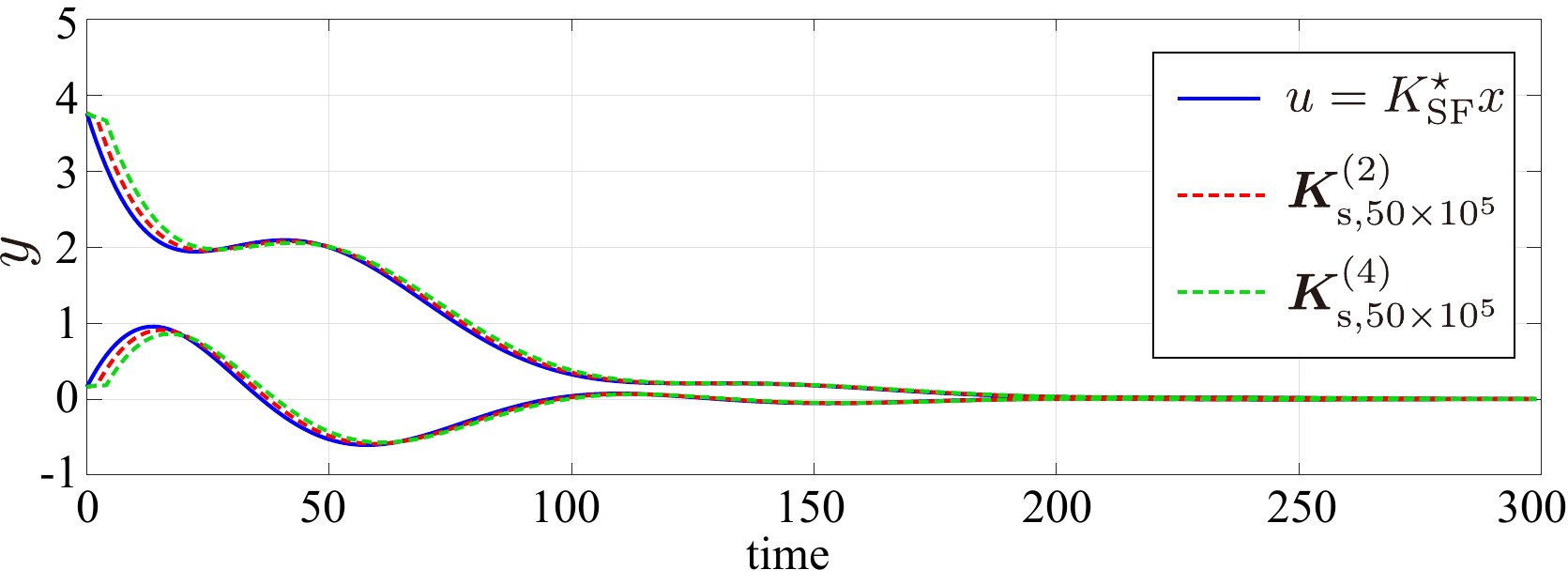}\vspace{-2mm}
    \caption{The blue solid line and red and green dotted lines show the trajectories of $y \in \mathbb R^2$ in \eqref{1} when 
    $u = K_{\rm SF}^{\star}x$, ${\bm K}_{{\rm s},50\times 10^5}^{(2)}$ and ${\bm K}_{{\rm s},50\times 10^5}^{(4)}$ are actuated at $t=0$, $t=2$, and $t=4$, respectively. } 
    \label{fig_perf_MB} 
  \end{center}\vspace{-4mm}
\end{figure}

We investigate the effectiveness of the proposed PGMs through a numerical example. Consider ${\bf \Sigma}_{\rm s}$ in \eqref{1} with
\begin{align*}
    & \hspace{-9mm} A \hspace{-0.5mm}=\hspace{-0.5mm} \begin{bmatrix}
 1.017 &  -0.0613 & 0.0281\\
    -0.0287 & 1.0471 & -0.0537\\
    -0.0564 & 0.1855 & 0.9256 
\end{bmatrix}, \quad B \hspace{-0.5mm}=\hspace{-0.5mm}
\begin{bmatrix} 
-0.0026\\
0.0019\\
0.0057
\end{bmatrix} \\
& \hspace{-9mm} C \hspace{-0.5mm}=\hspace{-0.5mm} 
\begin{bmatrix}    
  0 & 0.1201 & 0.6325 \\
  2.5992 & 0.4569 & 0 
\end{bmatrix}
\end{align*}
and $x(0) \sim \mathcal N(0, I_n)$, where $a \sim \mathcal N(\mu, V)$ denotes that $a$ follows a Gaussian distribution with mean $\mu$ and covariance $V$. This model is generated by the \texttt{drss} command in MATLAB. 
Let $Q=100I_r$ and $R=I_m$ in \eqref{defJ}. We take $L = 2$, which satisfies \eqref{condition L}. By applying $[u]^{0}_{L-1} \sim \mathcal N(0, I_{Lm})$ such that $\mathbb E[u(t)x^{\top}(0)] = 0$ for $t \in [0,L-1]$ to ${\bm \Sigma}_{\rm s}$, it follows that $\Phi = \mathcal P\mathcal P^{\top}$, where $\mathcal P$ is defined in \eqref{del_calP}. Hence, \eqref{def_psi} holds. 

We first investigate the global convergence of the model-based PGM \eqref{gd} with \eqref{8}. We choose $\alpha = 2.5\times 10^{-6}$ in \eqref{gd}. In Figure~\ref{fig_J_MB}, the blue solid line shows the variation of the cost $J(K_i)$ in \eqref{defJ} for the iteration $i$, while the black dotted line shows the optimal cost $J(K^{\star})$, where $K^{\star}$ is given by \eqref{optK}. In addition, for each of the updated gains indicated by the colored circles in Figure~\ref{fig_J_MB}, we construct ${\bm K}_{\rm s}$ in \eqref{dyn_K} based on Lemma \ref{lem2}. Similarly, we construct ${\bm K}_{\rm s}^{\star}$ from $K^{\star}$. Figure~\ref{fig_Ks_MB} shows the variation of the Bode diagrams of those controllers. We can see from Figures~\ref{fig_J_MB}-\ref{fig_Ks_MB} that the PGM \eqref{gd} with \eqref{8} successfully updates the dynamical controller so that its control performance is close to optimal. The resultant output feedback controller indicated by the red dotted line in Figure~\ref{fig_Ks_MB} is 
\begin{equation*}
 \Xi \hspace{-0.5mm}=\hspace{-0.5mm} \begin{bmatrix}
  0 & 0.2774\\
  1 & -0.04569
\end{bmatrix}\hspace{-1mm},~
\Lambda \hspace{-0.5mm}=\hspace{-0.5mm}
\begin{bmatrix} 
   28.73 & -29.47\\
   -32.98 & 37.41
\end{bmatrix}\hspace{-1mm},~
\Omega \hspace{-0.5mm}=\hspace{-0.5mm} 
\begin{bmatrix}    
    0 \\ 1
\end{bmatrix}^{\top}
\end{equation*}
and $\xi(0) = [137.5694, -0.6071]^{\top}$. Let this controller be denoted as ${\bm K}_{{\rm s}, 50 \times 10^5}^{(2)}$. In Figure~\ref{fig_perf_MB}, the red dotted lines show the trajectory of $y$ in \eqref{1} when ${\bm K}_{{\rm s}, 50 \times 10^5}^{(2)}$ is actuated at $t = 2$, whereas the blue solid lines show the case when the state-feedback controller $u = K^{\star}_{\rm SF}x$ is actuated at $t=0$, where $K^{\star}_{\rm SF}$ is defined as in \eqref{opt_SFB}. Note that $x(2)$ for the former case is generated by the aforementioned $x(0)$ and the initial input $[u]^0_{L-1}$. By comparing these lines, we can see that the learned output feedback dynamical controller exhibits an almost optimal performance equivalent to that of the state-feedback controller. 

\begin{figure}[t]
  \begin{center}
    \includegraphics[width=8.5cm]{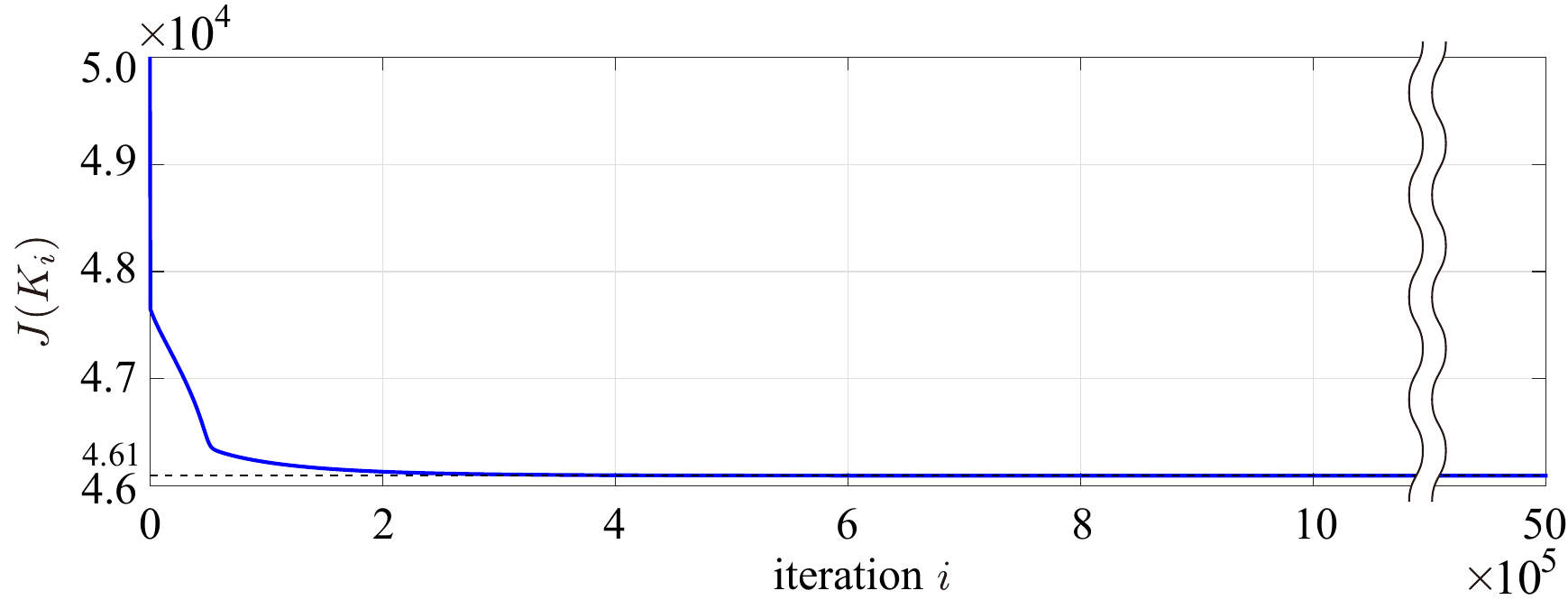} 
    \vspace{-4mm}
    \caption{Variation of $J(K_i)$ in \eqref{defJ} when $L=4$. } 
    \label{fig_J_MB_high} 
  \end{center}\vspace{-4mm}
\end{figure}
\begin{figure}[t]
  \begin{center} 
    \includegraphics[width=8.5cm]{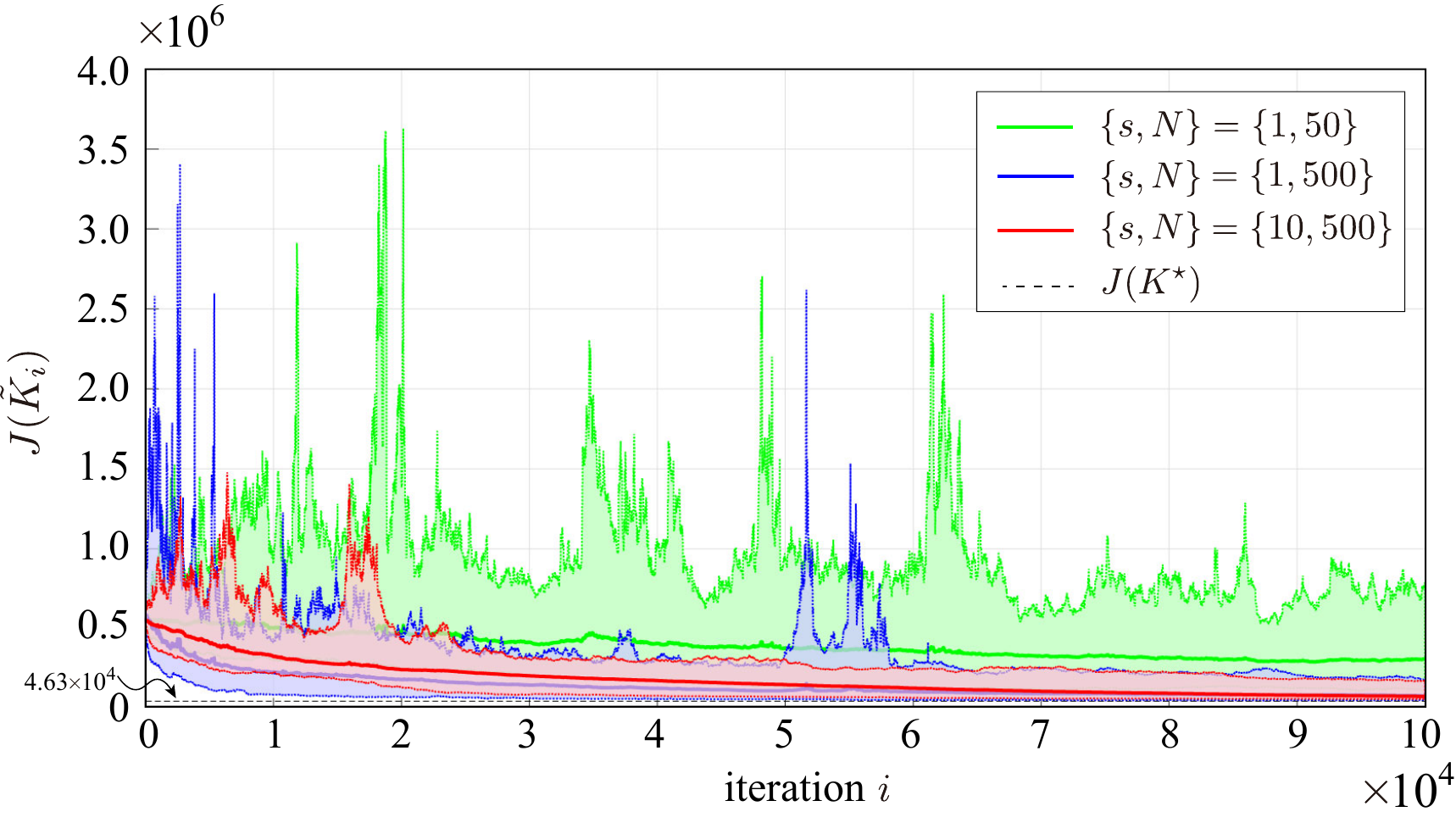} \vspace{-2mm}
    \caption{(Colored area) 50 variations of $J(\tilde{K}_i)$ in \eqref{defJ} for $\{s, N\} = \{1,50\}, \{1,500\}, \{10,500\}$ when $L=2$, where $\tilde{K}_i$ is generated by {\bf Algorithm 1}. (Colored broken lines) The average of the corresponding 50 variations.} 
    \label{fig_gainnorm}
  \end{center}\vspace{-4mm}
\end{figure}

Moreover, the global convergence property of the PGM holds even if we choose $L$ to be unnecessarily large. Let $L=4$ and the other settings be the same as above. Figure~\ref{fig_J_MB_high} shows the variation of $J$ in this case. Let the dynamical controller corresponding to the learned gain at $i=50 \times 10^5$ be denoted as ${\bm K}_{{\rm s},50\times 10^5}^{(4)}$. In Figure~\ref{fig_perf_MB}, the green dotted lines show the trajectory of $y$ when this controller is actuated at $t = 4$. We can see from Figures~\ref{fig_perf_MB}-\ref{fig_J_MB_high} that the PGM with a different choice of $L$ can also find an almost global optimal solution. 

\begin{figure}[t]
  \begin{center} 
    \includegraphics[width=8.5cm]{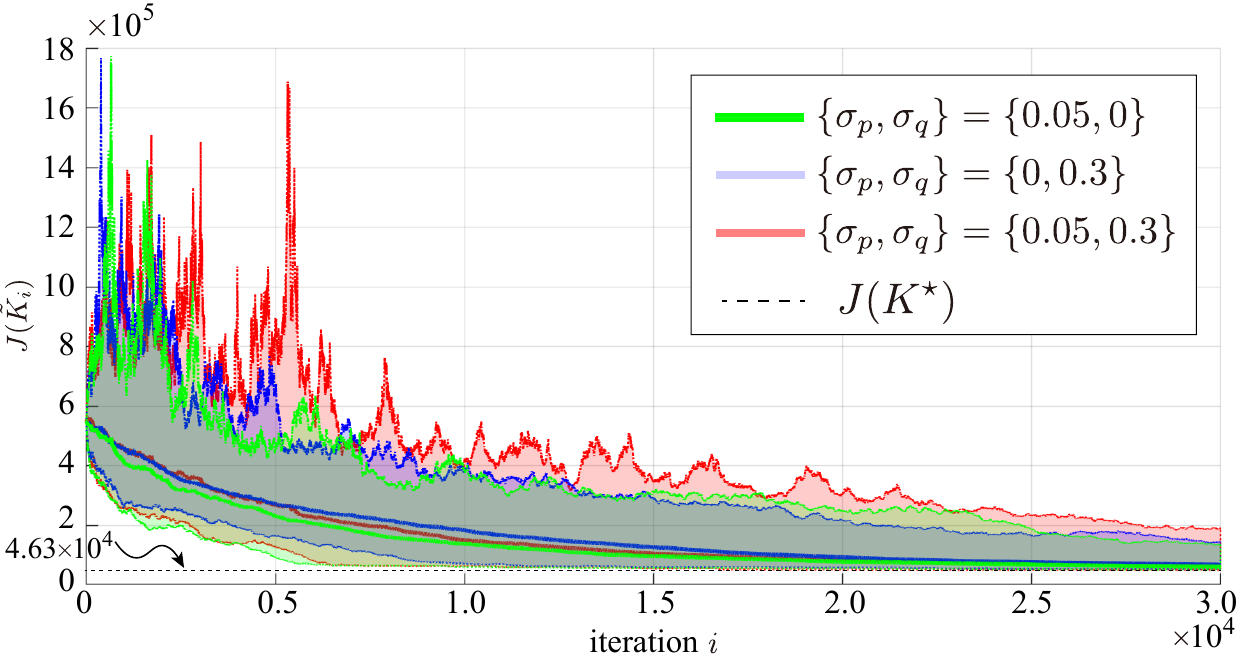} 
    \vspace{-4mm}
    \caption{(Colored area) 50 variations of $J(\tilde{K}_i)$ for $\{\sigma_p, \sigma_q\} = \{0.05, 0\}, \{0, 0.3\}, \{0.05, 0.3\}$, where the input-output data follow ${\bm \Sigma}_{\rm s}^{\rm noisy}$ in \eqref{sys_noisy}. (Colored broken lines) The average of the corresponding 50 variations.}
    \label{fig_noisycase}
  \end{center}\vspace{-4mm}
\end{figure}

\begin{figure}[t]
  \begin{center}
    \includegraphics[width=8.5cm]{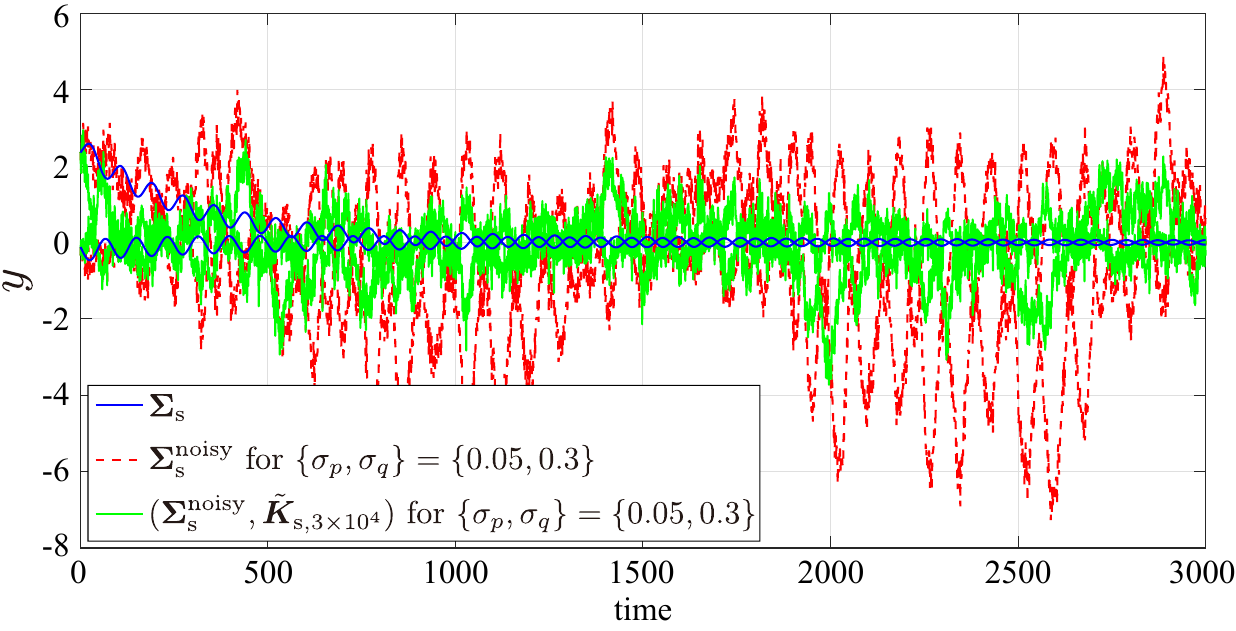} 
    \vspace{-4mm}
    \caption{Trajectory of $y$ of ${\bf \Sigma}_{\rm s}$ in \eqref{1}, that of ${\bf \Sigma}_{\rm s}^{\rm noisy}$ in \eqref{sys_noisy} for $\{\sigma_p, \sigma_q\} = \{0.05, 0.3\}$, and that of the closed-loop $({\bf \Sigma}_{\rm s}^{\rm noisy}, \tilde{\bm K}_{{\rm s}, 3\times 10^4})$, where $\tilde{\bm K}_{{\rm s}, 3\times 10^4}$ is constructed by \eqref{dyn_K} and \eqref{ABCD_hat} for learned gain $\tilde{K}_{3\times 10^4}$. }
    \label{fig_noisy_y} 
  \end{center}\vspace{-4mm}
\end{figure}

Next, we investigate the effectiveness of {\bf Algorithm 1}. Let $L=2$, $\alpha = 5\times 10^{-10}$ and $\delta = 0.05$. We consider three cases: $\{s, N\} = \{1,50\}, \{1,500\}, \{10,500\}$, and execute the algorithm 50 times for each case. In Figure \ref{fig_gainnorm}, the green, blue, and red areas show the ranges covered by the 50 different variations of $J$ for each case, while the broken colored lines are the averages of the corresponding 50 variations. In addition, the broken black line shows the optimal cost. This figure implies that the larger $s$ or $N$ is, the smaller the difference in performance improvement per iteration. Further, we can see that the controller adaptation to an optimal controller is successful for any choices of $s$ and $N$ as iterations continue.

Next, we investigate the effectiveness of {\bf Algorithm 1} in the presence of noise in the data. For this purpose, we consider 
\begin{equation}\label{sys_noisy}
{\bm \Sigma}_{\rm s}^{\rm noisy} : \hspace{0pt}
\begin{cases}
&x(t+1) = Ax(t) + Bu(t) + p(t) \\
&y(t) = Cx(t) + q(t)
\end{cases}, 
\end{equation}
where $p \sim \mathcal N(0, \sigma_p^2 I_n)$ represents the process noise, and $q \sim \mathcal N(0, \sigma_q^2 I_r)$ represents the observation noise. Instead of using the noiseless data following ${\bf \Sigma}_{\rm s}$ in \eqref{1}, we use the input-output data of this noisy system to learn the controller using {\bf Algorithm 1}.

Let $\{s, N\}=\{10, 200\}$, and keep the other parameters as well as $A$, $B$, and $C$, the same as above. We consider three cases: $\{\sigma_p, \sigma_q\}=\{0.05, 0\}, \{0, 0.3\}, \{0.05, 0.3\}$, and execute the algorithm 50 times for each case. To ensure a fair comparison among these cases, we quantify the control performance of the learned controller as $J$ in \eqref{defJ} with a common $\Phi$ for the closed-loop of the noiseless system ${\bf \Sigma}_{\rm s}$ and the learned controller. 
To observe the level of noise, we show a typical response of ${\bf \Sigma}_{\rm s}^{\rm noise}$ for $\{\sigma_p, \sigma_q\} = \{0.05, 0.3\}$ by the red dotted lines in Figure \ref{fig_noisy_y}, where a unit impulse input $u$ is applied and $x(0)$ is given such that $x(0) \sim \mathcal N(0, I_n)$. For comparison, we show the response of ${\bf \Sigma}_{\rm s}$ for the same input $u$ and $x(0)$ by the blue solid lines. We can see that the behavior of ${\bf \Sigma}_{\rm s}^{\rm noise}$ when $\{\sigma_p, \sigma_q\} = \{0.05, 0.3\}$ is significantly different from that of ${\bf \Sigma}_{\rm s}$ due to the noise.
In Figure~\ref{fig_noisycase}, the green, blue, and red areas represent the ranges covered by the 50 different variations of $J$ for each of the above three cases, while the solid-colored lines indicate the averages of the corresponding 50 variations. Additionally, the dashed black line represents $J(K^{\star})$, which is the same as in Fig~\ref{fig_gainnorm}. From this figure, we can see that the controller adaptation to $K^{\star}$ is successful even in situations where only process noise, only observation noise, or both are present. 
Moreover, the green lines in Figure~\ref{fig_noisy_y} depict the response of the closed-loop of ${\bm \Sigma}_{\rm s}^{\rm noisy}$ and the dynamic output feedback controller constructed by \eqref{dyn_K} and \eqref{ABCD_hat} for a learned gain $\tilde{K}_{3\times 10^4}$ that achieves $J(\tilde{K}_{3\times 10^4}) = 5.43 \times 10^4$. By comparing the red and green lines, we can see that the control performance has improved by the learned dynamic controller in the presence of noise, implying that the proposed algorithm can robustly learn dynamic output feedback controllers against noise. A theoretical analysis of this robust and globally convergent behavior is one of the important tasks for future research.

\begin{figure}[t]
  \begin{center}
    \includegraphics[width=8.5cm]{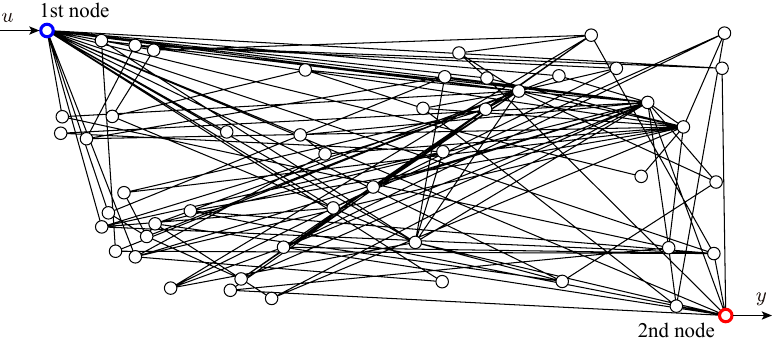} 
    \vspace{-4mm}
    \caption{Schematic diagram of the 50-dimensional network in \eqref{dyn_net}. }
    \label{graph_50}
  \end{center}\vspace{-4mm}
\end{figure}

\begin{figure}[t]
  \begin{center}
    \includegraphics[width=8.5cm]{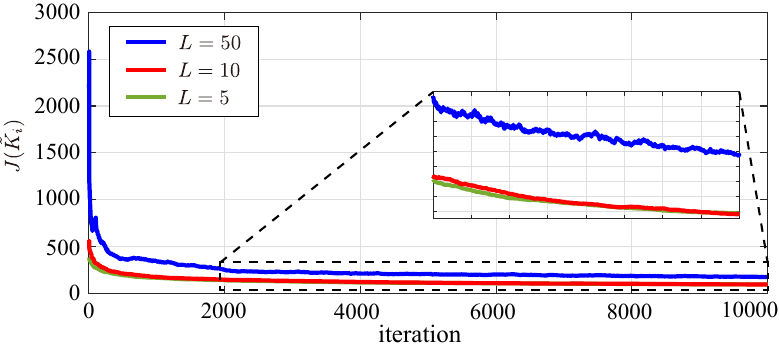} 
    \vspace{-4mm}
    \caption{The variation of $J(\tilde{K}_i)$ for $L=50, 10, 5$. }
    \label{investL}
  \end{center}\vspace{-4mm}
\end{figure}

\begin{figure}[t]
  \begin{center}
    \includegraphics[width=8.5cm]{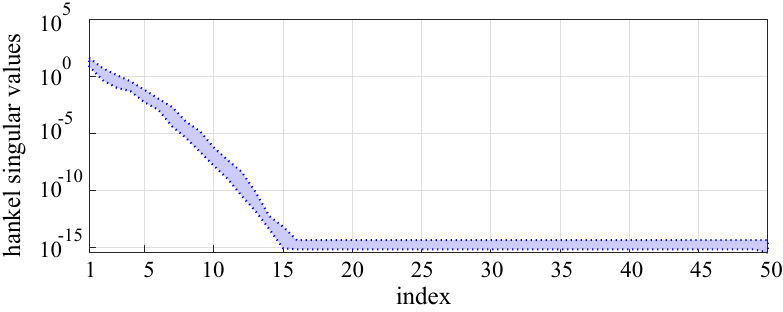} 
    \vspace{-4mm}
    \caption{50 variations of the hankel singular values of $(A, [B, x(0)], C)$-triple where $x(0)$ is randomly chosen from $\mathcal N(0, I)$. }
    \label{HSVDs}
  \end{center}\vspace{-4mm}
\end{figure}

Finally, we investigate the effectiveness of Algorithm 1 to large-scale systems. Let us consider a SISO network system composed of $50$ nodes. For $j\in \{1,\ldots,50\}$, the $j$-th node dynamics is described as follows: 
\begin{align}\label{dyn_net}
\hspace{-8mm}
\begin{cases}
    &\hspace{-3mm}x_j(t+1) = a_j x_j(t) + \sum_{k \in \mathcal N_j} a_{jk}(x_k(t)-x_j(t)) + b_ju(t)\\
    &\hspace{-3mm}y(t) = c_j x_j(t)
\end{cases} 
\end{align}
where $x_j \in \mathbb R$ is the state of the $j$-th node, $\mathcal N_j$ is the set of nodes connecting to the $j$-th node, and $a_{jk}$ is randomly chosen from the range $(0,0.089]$ if $k \in \mathcal N_j$, otherwise $a_{jk} = 0$, and $a_j \coloneqq -\sum_{k \in \mathcal N_j} a_{jk}$. Assume the input is applied to the first node with its weight one, whereas the output is the state of the second node, i.e., $b_1 = c_2 = 1$ and $b_j=c_j=0$ for other $j$. The interconnection structure among nodes is an undirected graph generated by the Barab\'{a}si–Albert model \cite{albert2002statistical} with its parameter set to three; see Figure \ref{graph_50} for details of this network diagram. This network can be expressed in the form of equation \eqref{1} for $n=50$, and is stable. Let $x(0) \sim \mathcal N(0, I_n)$ and $[u]_{L-1}^0 \sim \mathcal N(0, I_{Lm})$. 

In the following, we consider three cases, namely $L=50$, $10$, and $5$, and apply {\bf Algorithm 1} to the network system for each case. All these cases do not satisfy the condition \eqref{condition L} because $n$ is large. Therefore, Lemma \ref{lem_VARX} does not hold, as a result of which, the model-based PGM \eqref{gd} with \eqref{8} is not executable. However, even in this situation, {\bf Algorithm 1} is executable and the IOH-feedback controller \eqref{IOH_law} can be equivalently converted to ${\bm K}_{\rm s}$ in \eqref{dyn_K}. Therefore, we can learn a dynamic controller in the form of \eqref{dyn_K} by {\bf Algorithm 1}. Because the condition \eqref{condition L} is not satisfied, for calculating the cost in \eqref{defJ}, $\{u, y\}$ is assumed to obey the closed-loop $({\bm \Sigma}_{\rm s}, {\bm K}_{{\rm s},i})$, where ${\bm K}_{{\rm s},i}$ is the converted dynamic controller. 

Let $s=10$, $N=500$, $\alpha=10^{-8}$, $K_0=0$ in {\bf Algorithm 1}. Figure \ref{investL} shows the variation of the costs for the three cases. The blue line shows that the proposed algorithm can improve control performance even when $L$ is large such as $L=50$. Furthermore, as depicted by the red and green lines, even when $L$ is chosen {\it small}, the algorithm can successfully improve control performance. To investigate the cause of this performance improvement, we generate $50$ random instances of $x(0)$ from $\mathcal N(0, I_n)$ and compute the Hankel singular values $\rho_1 \geq \ldots \geq \rho_{50}$ of the $(A, [B, x(0)], C)$-triple for each of the $50$ cases. In Figure \ref{HSVDs}, the blue area represents the region bounded by the maximum and minimum values of $50$ values of $\rho_i$ for each $i$. This figure shows that, in terms of input-output characteristics including the initial state disturbance, the network system can be approximated as a low-dimensional system such as a 10-dimensional system. Because input-output data is used in the proposed algorithm, if the input-output characteristics can be approximated  as a low-dimensional system, the learning to the original system can be regarded to be that to a low-dimensional system, thereby successfully learning the controller even in the case of $L=10$. Although presenting a data-driven guideline for choosing $L$ would be difficult, the success in learning for a very small value of $L$ such as $L=5$, suggests that such an {\it appropriate} choice of $L$ is not necessary. 
Finally, we investigate the computational cost of {\bf Algorithm 1} with respect to $L$. Because Step 2) of {\bf Algorithm 1} is for computing the time evolution of the system, we define $\tau_L$ as the computational time for executing Steps 1), 3)-6), excluding Step 2). In the three cases, $\tau_5=0.9, \tau_{10}=1.1, \tau_{50}= 2.5$ [msec], where all codes were developed in 2020a and run in an Intel(R) Core(TM) i7-10875H 2.30GHz, RAM 64.0GB computer. This result implies that the growth in computation time with respect to $L$ is negligible.

\section{Conclusion}\label{conclusion}
In this paper, we have proposed model-based and model-free PGMs for designing dynamic output feedback controllers for discrete-time partially observable systems, using the newly introduced IOH dynamics and the corresponding reachability-based lossless projection. Unlike in SOF control \cite{fatkhullin2021optimizing,takakura2022structured,bu2019topological,bu2019lqr}, the global linear convergence of the PGMs is guaranteed with a high probability (which becomes one for the model-based PGM) when design parameters are appropriately chosen.
Moreover, we have numerically demonstrated that the proposed algorithm can successfully learn nearly optimal dynamic output feedback controllers even when input-output data are contaminated by noise. Future work includes the analysis of this robust and global convergence for noisy cases and the development of a systematic method of finding initial stabilizing controllers from the data, and to apply the proposed method to a realistic systems such as power systems \cite{sadamoto2019dynamic}. 

\bibliographystyle{elsarticle-num}
\bibliography{refs}

\appendix

\section{Proof of Lemma \ref{lem_VARX}}\label{pf_App1}
Note that \eqref{1} yields
\begin{eqnarray}
x(t) &\hspace{-2mm}=&\hspace{-2mm} A^Lx(t-L) + \mathcal R_L({\bm \Sigma}_{\rm s})[u]_{t-1}^{t-L}, \label{sys_dyn_y1} \\
{[y]}_{t-1}^{t-L} &\hspace{-2mm}=&\hspace{-2mm} \mathcal O_L({\bm \Sigma}_{\rm s})x(t-L) + \mathcal H_L({\bm \Sigma}_{\rm s})[u]_{t-1}^{t-L}. \label{sys_dyn_y} 
\end{eqnarray}
From \eqref{sys_dyn_y}, $x(t-L) = \mathcal O_L^{\dagger}({\bm \Sigma}_{\rm s})({[y]}_{t-1}^{t-L} - \mathcal H_L({\bm \Sigma}_{\rm s})[u]_{t-1}^{t-L})$ follows. By substituting this into \eqref{sys_dyn_y1}, we have
\begin{equation}\label{xisGammav}
    x(t) = \Gamma v(t)
\end{equation}
for any $t \geq L$. Hence, from the output equation in \eqref{1}, we have the output equation in \eqref{dyn_IOH}. 
The dynamics of \eqref{dyn_IOH} follows from the definition of $v$. This shows that $v$ and $y$ obey \eqref{dyn_IOH}. Next, we show \eqref{init_v}. 
From \eqref{sys_dyn_y}, any $v(t)$ satisfies
\begin{equation}\label{del_calP}
    v(t) = \mathcal P \left[\hspace{-0.5mm}\begin{array}{c} [u]^{t-L}_{t-1} \\ x(t-L) \end{array}\hspace{-0.5mm}\right]\hspace{-0.5mm},~ \mathcal P \coloneqq \left[\hspace{-0.5mm}\begin{array}{cc}I & \\ \mathcal H_L({\bm \Sigma}_{\rm s}) & \mathcal O_L({\bm \Sigma}_{\rm s}) \end{array}\hspace{-0.5mm}\right] 
\end{equation}
for $t \geq L$. From the reachability of ${\bm \Sigma}_{\rm s}$, the subspace spanned by $v(t)$ for any $t \geq L$ and input signals is $\im \mathcal P$. Note that this holds without depending on $[u]^{0}_{L-1}$ and $x(0)$. Hence, 
\begin{equation}
\label{ppp}
 \im \mathcal P = \im [\Theta^{L(m+r)-1}[\Pi, v(L)], \ldots, [\Pi, v(L)]] = \mathscr{P}, 
\end{equation}
which yields \eqref{init_v}. Note that $\rank \mathcal P = Lm+n$. Hence, ${\rm dim}\mathscr{P} = Lm+n$. This completes the proof. \qed

\section{Proof of Lemma \ref{lem2}}\label{pf_App2}
We first show the equivalence of the two open-loop systems ${\bm K}$ and ${\bm K}_{\rm s}$ for a common input $y$. Let $u$ in \eqref{IOH_law} be denoted by $\check{u}$, and the IOH generated by \eqref{IOH_law} be denoted by $\check{v}(t) \coloneqq [([\check{u}]^{t-L}_{t-1})^{\top}, ({[y]}^{t-L}_{t-1})^{\top}]^{\top}$ for $t \geq L$. Hence, \eqref{IOH_law} is described as $\check{u}(t) = K\check{v}(t)$ for $t \geq L$, and $\xi(0)$ is as $\xi(0) = \mathcal O_L^{-1}({\bm K}_{\rm s})[I, -\mathcal H_L({\bm K}_{\rm s})] \check{v}(L)$. We show $\check{u}(t) = u(t)$ for $t \geq 0$. From \eqref{dyn_K}, we have 
${[u]}_{L-1}^{0} = \mathcal O_L({\bm K}_{\rm s})\xi(0) + \mathcal H_L({\bm K}_{\rm s})[y]_{L-1}^{0}. 
$
By substituting $\xi(0)$ into this, we have $[u - \check{u}]_{L-1}^{0} = 0$. Hence, $v(L) = \check{v}(L)$. Further, similarly to the proof of Lemma \ref{lem_VARX}, it follows from \eqref{dyn_K} that ${u}(t) = Kv(t)$ for $t \geq L$. Therefore, $\check{u}(t) = u(t)$ for $t \geq 0$, which shows the equivalence between ${\bm K}$ and ${\bm K}_{\rm s}$. Further, ${\bm \Sigma}$ and ${\bm \Sigma}_{\rm s}$ are equivalent from Lemma \ref{lem_VARX}. This completes the proof. \qed

\section{Proof of Proposition \ref{prop0_1}} \label{app_propo}
We first show 
\begin{equation}\label{app_propo_gam_1}
    x(t) = (\Gamma + \bar{\Gamma})v(t), \quad t \geq L. 
\end{equation}
Note that \eqref{xisGammav} holds for $t \geq L$. Further, because $\bar{\Gamma} v = 0$ due to ${\rm ker} \bar{\Gamma} = \mathscr{P}$ and \eqref{init_v} follows, \eqref{app_propo_gam_1} holds. Based on this relation, we show the claim. 

To this end, we define the value function described as 
\begin{equation}\label{app_propo_eq_2}
    V \hspace{-0.5mm}\coloneqq\hspace{-1mm} \underset{u(t), t \geq L}{\rm min}\hspace{3pt}{\mathbb E}_{v(L)\sim{\mathcal N}_{\Phi}}\hspace{-1mm}\left[\sum_{t = L}^{\infty}y^{\top}(t)Qy(t) + u^{\top}(t)Ru(t)\hspace{-0.5mm}\right],
\end{equation}
where $y$ obeys \eqref{dyn_IOH}. 
From Lemma~\ref{lem_VARX}, without loss of generality, $y(t)$ in \eqref{app_propo_eq_2} can be assumed to obey \eqref{1}. Moreover, from \eqref{app_propo_gam_1} and \eqref{def_psi}, the fact $v(L) \sim \mathcal N_{\Phi}$ yields $x(L) \sim \mathcal N_{\Phi_x}$ where $\Phi_x \coloneqq \Gamma \Phi \Gamma^{\top}$. Hence, \eqref{app_propo_eq_2} can be written as 
\begin{equation}\label{app_propo_eq_4_}
    V \hspace{-0.5mm}=\hspace{-1mm} \underset{u(t), t \geq L}{\rm min}\hspace{3pt}{\mathbb E}_{x(L)\sim{\mathcal N}_{\Phi_x}}\hspace{-1mm}\left[\sum_{t = L}^{\infty}y^{\top}(t)Qy(t) + u^{\top}(t)Ru(t)\hspace{-0.5mm}\right]\hspace{-1mm},
\end{equation}
where $y$ obeys \eqref{1}. From optimal control theory \cite{lewis2012optimal}, if $\Phi_x > 0$, the function $V$ is written as $V = \mathbb E[x^{\top}(L)Xx(L)]$ where $X$ is a solution to ${\rm Ric}(X) = 0$, while the minimizer $u(\cdot)$ in \eqref{app_propo_eq_4_} is described as 
\begin{align}\label{prop1_pf1}
u(t) = K_{\rm SF}^{\star}x(t), \quad t \geq L,   
\end{align}
where $K_{\rm SF}^{\star}$ is defined as \eqref{opt_SFB}. By combining \eqref{prop1_pf1} with \eqref{app_propo_gam_1}, we have \eqref{optK}. Therefore, showing $\Phi_x > 0$ should suffice. From \eqref{def_psi} and \eqref{prop_P}, there exists $\hat{\Phi} > 0$ such that $\Phi = P\hat{\Phi}P^{\top}$. From \eqref{dyn_IOH2} and \eqref{xisGammav}, it follows that $x(t) = \Gamma P \hat{v}(t)$ for $t \geq L$. Because ${\bm \Sigma}_{\rm s}$ is reachable, for any $\bar{x} \in \mathbb R^{n}$, an input sequence exists to reach $\bar{x}$; therefore, there exists $\bar{\hat{v}}$ such that $\bar{x} = \Gamma P \bar{\hat{v}}$. To satisfy this for any $\bar{x} \in \mathbb R^{n}$, $\Gamma P$ must satisfy $\rank \Gamma P = n$. Therefore, $\Phi_x = (\Gamma P)\hat{\Phi} (\Gamma P)^{\top}$ is invertible because $\hat{\Phi}$ is invertible. This completes the proof. 
\qed

\section{Proof of Lemma \ref{lem_DYC}}\label{pf_App3}
The existence of $P$ satisfying \eqref{prop_P} follows from Lemma \ref{lem_VARX}. Next, we show \eqref{dyn_IOH2}. Define $\overline{P} \in {\mathbb R}^{(Lm+Lr )\times (Lr-n)}$ such that $[P, \overline{P}]$ is a unitary matrix. Define $\hat{v} \coloneqq P^{\top}v$ and $\overline{v} \coloneqq \overline{P}^{\top}v$. Then, the dynamics in \eqref{dyn_IOH} can be written as
\begin{equation*}\label{control z}
    \begin{bmatrix}
         \hat{v}(t+1)\\
         \overline{v}(t+1)
    \end{bmatrix}
    = 
    \begin{bmatrix}
         P^{\top}\Theta P & P^{\top}\Theta \overline{P}\\
         \overline{P}^{\top}\Theta P & \overline{P}^{\top}\Theta \overline{P}
    \end{bmatrix}
    \begin{bmatrix}
         \hat{v}(t)\\
         \overline{v}(t)
    \end{bmatrix}
    + 
    \begin{bmatrix}
         P^{\top}\Pi\\
         \overline{P}^{\top}\Pi
    \end{bmatrix}
    u(t). 
\end{equation*}
It follows from \eqref{init_v} that $\overline{v} \equiv 0$. Hence, the dynamics in \eqref{dyn_IOH2} follows. Finally, we have $v = P\hat{v} + \overline{P}\overline{v} = P\hat{v}$. This shows \eqref{dyn_IOH2}. From \eqref{dyn_IOH2}, we have $\hat{v}(L) = P^{\top}v(L)$. Hence, $\hat{v}(L) \sim \mathcal N_{\hat{\Phi}}$. The positive definiteness of $\hat{\Phi}$ immediately follows from \eqref{def_psi}. This completes the proof.  \qed

\section{Proof of Lemma~\ref{gradient}} \label{app_lemcc}
We define 
\begin{equation}
\label{V}
    C_{K_i}(v(t)) \coloneqq v^{\top}(t)\Psi_{K_i} v(t).
\end{equation}
By pre- and post-multiplying $P$ and $P^{\top}$ by \eqref{5}, and by using the relation \eqref{def_PSI}, we have $
\Psi_{K_i} = PP^{\top}(\Theta_{K_i}^{\top}\Psi_{K_i}\Theta_{K_i} + \Gamma^{\top}C^{\top}QC \Gamma + K_i^{\top}RK_i)PP^{\top}$. Note that $PP^{\top}v = v$ holds for any $v$ because of \eqref{init_v} and \eqref{prop_P}. By substituting the above $\Psi_{K_i}$ to $C_{K_i}$, $C_{K_i}$ at $t=L$ can be written as $ C_{K_i}(v(L)) \hspace{-0.5mm}=\hspace{-0.5mm} v^{\top}(L)(\Gamma^{\top}C^{\top}QC\Gamma \hspace{-0.5mm}+\hspace{-0.5mm} K_i^{\top}RK_i)v(L) \hspace{-0.5mm}+\hspace{-0.5mm} C_{K_i}(v(L+1))$. 
Similarly to Lemma 1 in \cite{fazel}, by differentiating $C_{K_i}$ at $K_i$ and taking the expectation of the result, the claim follows. \qed

\section{Proof of Lemma~\ref{lem_stab}}\label{app_lem_stab}
The sufficiency is obvious. We show the necessity. 
 Note that
 \[
 \scalebox{0.95}{$\displaystyle 
 \textstyle \sum_{t=(k-1)L}^{kL-1}\left(y^{\top}(t)Qy(t) + u^{\top}(t)Ru(t)\right) = v^{\top}(kL)G v(kL) $}
 \]
  holds for $k = 2,3,\cdots$, where 
 \begin{align}\label{defGss}
 G \coloneqq \left[
 \begin{array}{cc}
 I_L \otimes R & \\ & I_L \otimes Q
 \end{array}
 \right] > 0
 \end{align}
 and $\otimes$ is the Kronecker product. Hence, we have $J(K) = \mathbb E_{\hat{v}(L)\sim \mathcal N_{\hat{\Phi}}}\left[\sum_{k=2}^{\infty} \hat{v}^{\top}(kL)\hat{G}\hat{v}(kL)\right]$, where 
 \begin{align}\label{defGhat}
 \hat{G} \coloneqq P^{\top}GP > 0,    
 \end{align}
 and $\hat{\Phi}$ is defined in \eqref{def_Phihat}. Further, using the notation 
 \begin{equation}\label{def_Omega_l}
  \textstyle  \Omega_l \coloneqq \mathbb E_{\hat{v}(L)\sim \mathcal N_{\hat{\Phi}}}\left[\sum_{k=2}^{\infty} \hat{v}(kL+l)\hat{v}^{\top}(kL+l)\right] 
 \end{equation}
 for given $l = 0,\ldots, L-1$, the cost $J$ can be written as $J(K)= \tr(\hat{G}\Omega_0)$. We now use the following formula: For $A \geq 0$ and $B \geq 0$, we have
\begin{align}\label{formula11}
\tr(AB) \geq \sigma_{\rm min}(A) \tr(B). 
\end{align}
This can be shown as follows. Since $B \geq 0$, there exists $B^{1/2}$ such that $B^{\top/2} = B^{1/2} \geq 0$. Since $A \geq 0$, we have $A \geq \sigma_{\rm min}(A)I \Rightarrow B^{1/2}AB^{1/2} \geq \sigma_{\rm min}(A)B$. Therefore, \eqref{formula11} follows. By letting $A \leftarrow \hat{G}$ and $B \leftarrow \Omega_0$ in \eqref{formula11}, we have $J(K) \geq \sigma_{\rm min}(\hat{G}) \tr(\Omega_0)$. Hence, $J(K) < \infty$ yields $\tr(\Omega_0) < \infty$, yielding $\hat{v}(\infty) = 0$. This shows the stability of $\hat{\Theta}_K$. \qed

\section{Proof of Lemma \ref{PL Lem MIMO}}\label{SISO Case}
The invertibility of $\hat{V}_K$ follows from $\hat{V}_K \geq \hat{\Phi} > 0$, where $\hat{\Phi}$ is defined in \eqref{def_Phihat}. 
Consider $C_K(\cdot)$ in \eqref{V}, $E_K$ and $V_K$ in \eqref{def_EV}. Note that $v(t) = (\Theta + \Pi K)^{t-L}v(L)$ for $t \geq L$. Define $v_{\star}(t) = (\Theta + \Pi K^{\star})^{t-L}v(L)$, 
$c(t) \coloneqq v^{\top}(t)(\Gamma^{\top}C^{\top}QC\Gamma  + K^{\top}RK)v(t)$, and
$c_{\star}(t) \coloneqq v_{\star}^{\top}(t)(\Gamma^{\top}C^{\top}QC\Gamma  + K^{\top}RK)v_{\star}(t)$ for $t\geq L$. It follows from the above that
\begin{eqnarray}
\hspace{-6mm}&&C_{K}(v(L)) - C_{K^{\star}}(v(L)) \nonumber \\
\hspace{-6mm}&&= C_{K}(v(L)) - \textstyle \sum_{t=L}^{\infty} \left(c_{\star}(t) + C_{K}(v_{\star}(t)) - C_{K}(v_{\star}(t)) \right) \nonumber \\
\hspace{-6mm}&&= -\textstyle \sum_{t=L}^{\infty} \left(c_{\star}(t) + C_{K}(v_{\star}(t+1)) - C_{K}(v_{\star}(t)) \right) \nonumber
\end{eqnarray}
Hence, similarly to the proof of Lemma 11 in \cite{fazel}, we have
\begin{eqnarray}
\hspace{-12mm}&&J(K)-J(K^{\star}) \nonumber \\
\hspace{-12mm}&&\leq \textstyle\sum_{t=L}^{\infty} \tr\left(v_{\star}(t)v_{\star}^{\top}(t)E_K^{\top}(R+ \Pi^{\top}\Psi_K \Pi)^{-1}E_K\right) \label{SISO_prr2}\\
\hspace{-12mm}&& = \tr (V_{K^{\star}}E_K^{\top}(R + \Pi^{\top}\Psi_K\Pi)^{-1}E_K) \label{SISSS}
\end{eqnarray}
where appendix C.1 of \cite{fazel} is used for deriving \eqref{SISO_prr2}. 
Further, it follows from $\hat{v} = P^{\top}v$ that $V_K$ is written as 
\begin{equation}\label{hatSigma}
    V_K = P\hat{V}_KP^{\top}.
\end{equation}
Hence, $\nabla J$ in \eqref{8} can be written as 
\begin{equation}\label{re gradient}
    \nabla J(K)  = 2E_{K}V_{K} = 2\hat{E}_K\hat{V}_KP^{\top}, \quad \hat{E}_K \coloneqq E_KP. 
\end{equation}
Since $P^{\top}P = I$, it follows from the first equation that 
\begin{equation}\label{hatE_Y}
    \hat{E}_K = \frac{1}{2}\nabla J(K)P\hat{V}_K^{-1}.
\end{equation}
Using \eqref{hatSigma}-\eqref{re gradient} and similarly to the proof of Lemma 11 in \cite{fazel}, we can rewrite \eqref{SISSS} as 
\begin{eqnarray}
J(K) - J(K^{\star}) &\hspace{-2mm}\leq&\hspace{-2mm} \tr (\hat{V}_{K^{\star}}\hat{E}_K^{\top}(R + \Pi^{\top}\Psi_K\Pi)^{-1}\hat{E}_K) \nonumber \\ 
&\hspace{-2mm}\leq&\hspace{-2mm}  \frac{\|\hat{V}_{K^{\star}}\|}{\sigma_{\rm min}(R)}\tr (\hat{E}_K^{\top}\hat{E}_K). \label{fff}
\end{eqnarray}
Finally, by substituting \eqref{hatE_Y} into \eqref{fff} and using $\|P\| = 1$, the claim follows.\qed

\section{Proof of Lemma~\ref{smoothness_new}}\label{app_smoothness_new}
Similarly to deriving the final inequality in page 31 of \cite{bu2019lqr}, for any $K \in \mathbb K$ we have the operator 2-norm of the Hessian $\nabla^2J(K)$ as 
\begin{align}
   \hspace{-8mm}\|\nabla^2 J(K)\|\leq 2 \gamma \|\hat{V}_K\| &+ 4 \left\|\hat{\Theta}_K \hat{V}_K^{\frac{1}{2}} \right\| \nonumber \\ 
   &\times \underset{\|K'\|_F=1}{\sup} \left\|\hat{K}'^{\top}\hat{\Pi}^{\top}X\right\| {\rm tr}(\hat{V}_K^{\frac{1}{2}}) \label{l8_pf},
\end{align}
where $\gamma \coloneqq \|\hat{\Pi}^{\top}\hat{\Psi}_{K}\hat{\Pi} + R \|$, $\hat{\Pi}^{\top} \coloneqq P^{\top}\Pi$, $X \coloneqq (\partial \hat{\Psi}_{K_{\alpha'}} / \partial \alpha')_{\alpha' =0}$, $K_{\alpha'} \coloneqq K + \alpha'K'$, $\hat{V}_K$ is defined in Lemma~\ref{PL Lem MIMO} and $\Psi_K$ is defined in \eqref{def_PSI}. Because \eqref{def_qqq} is equivalent to $\|\nabla^2 J(K')\| \leq q$ \cite{boyd2004convex}, it suffices to show that the RHS of \eqref{l8_pf} is bounded by $q$. 

Note that $J(K) = \tr(\hat{\Phi}\hat{\Psi}_{K})$ from the first equation in \eqref{6} and \eqref{def_Phihat}. Hence, for any $K \in \mathbb K$, it follows that
\begin{equation} \label{bound Psi}
  \|\hat{\Psi}_{K}\|\leq \tr(\hat{\Psi}_{K})\leq J(K) / \sigma_{\min}(\hat{\Phi}) \leq c / \sigma_{\min}(\hat{\Phi}). 
\end{equation}
where the second inequality follows from \eqref{formula11} with $A \leftarrow \hat{\Phi}$ and $B \leftarrow \hat{\Psi}_K$, and the last inequality follows from \eqref{def_Kdom}. Using \eqref{bound Psi} and $\|\hat{\Pi}\| \leq 1$, we have 
\begin{align}\label{boundofgamma}
 \gamma \leq c/\sigma_{\min}(\hat{\Phi}) + \|R\|. 
\end{align}
We next show an upper bound of $\|\hat{V}_K\|$. 
Given $l = 0, \ldots, L-1$, consider $\Omega_l$ in \eqref{def_Omega_l} and $\hat{G}$ in \eqref{defGhat}. Similarly to the proof of Lemma~\ref{lem_stab}, we have $J(K) = \mathbb E[\sum_{j=0}^{l-1} y^{\top}(L+j)Qy(L+j) + u^{\top}(L+j)Ru(L+j)] + \tr(\hat{G}\Omega_l)$. Hence, $\tr(\hat{G}\Omega_l) \leq J(K) \leq c$ for any $K \in \mathbb K$. Note here that $\hat{V}_{K} = \sum_{l=0}^{L-1} \Omega_l$. Hence,
\begin{align}
    & \hspace{-6mm}\|\hat{V}_{K}\| = \|\textstyle \sum_{l=0}^{L-1} \Omega_l\| \leq \textstyle \sum_{l=0}^{L-1} \|\Omega_l\| \leq \textstyle \sum_{l=0}^{L-1} \tr (\Omega_l) \nonumber \\ 
    & \hspace{-6mm}\leq \textstyle \sum_{l=0}^{L-1} \frac{\tr (\hat{G}\Omega_l)}{\sigma_{\rm min}(\hat{G})} \leq \textstyle \sum_{l=0}^{L-1} \frac{c}{\sigma_{\rm min}(\hat{G})} = \frac{Lc}{\sigma_{\rm min}(\hat{G})}, 
\end{align}
where the third inequality follows from \eqref{formula11} with $A \leftarrow \hat{\Phi}$ and $B \leftarrow \Omega_l$. Further, from the definition of $\rho$ in \eqref{def_rhoQ} and $G$ in \eqref{defGss}, we have $G \geq \rho I$, which yields $\hat{G} \geq \rho I \iff \sigma_{\rm min}(\hat{G}) \geq \rho$. Hence, 
\begin{equation}\label{lm_rho}
\|\hat{V}_{K}\| \leq Lc / \rho    
\end{equation}
holds. We next show an upper bound of $\|\hat{\Theta}_K \hat{V}_K^{1/2}\|$. Note that $\hat{\Theta}_K\hat{V}_K\hat{\Theta}_K^{\top} \leq \hat{V}_K$ follows from the definition of $\hat{V}_K$ and that $\hat{\Theta}_K$ is Schur. Hence, 
\begin{align}
   \|\hat{\Theta}_K \hat{V}_K^{1/2}\| &= \left(\|\hat{\Theta}_K\hat{V}_K^{1/2}\|^2 \right)^{\frac{1}{2}} = \|\hat{\Theta}_K\hat{V}_K\hat{\Theta}_K^{\top}\|^{\frac{1}{2}} \nonumber \\
   & \leq \|\hat{V}_K\|^{1/2} \leq \left(Lc / \rho\right)^{1/2}. \label{boundofThV}
\end{align}
Furthermore, 
\begin{align}
\hspace{-8mm} {\rm tr}(\hat{V}_K^{\frac{1}{2}}) \leq \sqrt{2(Lm+n)}\|\hat{V}_K\|^{1/2} \leq \sqrt{2(Lm+n)Lc/\rho}      \label{boudnoftrace}
\end{align}
holds, where the first inequality follows from the proof of Proposition 7.7 in \cite{bu2019lqr}. 

For the supremum term in \eqref{l8_pf}, similarly to the last inequality in page 32 in \cite{bu2019lqr}, it follows that 
\[
X - \hat{\Theta}_K^{\top}X \hat{\Theta}_K \leq M
\]
where $M \coloneqq \hat{\Psi}_K - \hat{Q} + \hat{K}'^{\top}\hat{\Pi}^{\top}\hat{\Psi}_K\hat{\Pi}\hat{K}' + \hat{K}'^{\top}R\hat{K}'$. Thus, $\|X\| \leq \tr(X) \leq \tr(\sum_{l=0}^{\infty}(\hat{\Theta}_K^{\top})^lM \hat{\Theta}_K)$. Therefore, by using the facts $\sup_{\|A\|_F \leq 1}\|AB\| \leq \|B\|$ for any $A, B$ and $\|\hat{\Pi}\| \leq 1$, we have
\begin{equation}
  \scalebox{0.94}{$\displaystyle 
\underset{\|K'\|_F=1}{\sup} \left\|\hat{K}'^{\top}\hat{\Pi}^{\top}X\right\| \leq \underset{\|K'\|_F=1}{\sup} \tr\left(\sum_{l=0}^{\infty}(\hat{\Theta}_K^{\top})^lM\hat{\Theta}_K^l\right) \label{new_expa}. $}
\end{equation}
From simple calculation, it follows for any $K'$ satisfying $\|K'\|_F=1$ that 
\begin{align} \label{boundofM}
{\rm tr}(M) \leq 2c / \sigma_{\min }(\hat{\Phi}) + \tr(R)-\tr(\hat{Q}).     
\end{align}
Now we use the following formula: For $Y \geq 0$, $X > 0$, and a Schur matrix $A$, we have 
\begin{align}
&\hspace{-8mm}{\rm tr}\left(\textstyle \sum_{l=0}^{\infty} (A^{\top})^l Y A^l \right) 
= {\rm tr}\left(\textstyle \left(\sum_{l=0}^{\infty} (A^{\top})^l A^l\right)Y \right) \nonumber \\
&\hspace{-9mm}\leq \|\textstyle \sum_{l=0}^{\infty} A^l (A^{\top})^l\| {\rm tr}(Y) \leq \frac{\tr (Y)}{\sigma_{\rm min}(X)}\|\textstyle \sum_{l=0}^{\infty}A^lX(A^{\top})^l\| \label{form111}
\end{align}
where the last inequality follows from $I \leq X/\sigma_{\rm min}(X) \Rightarrow A^l(A^{\top})^l \leq A^lX(A^{\top})^l / \sigma_{\rm min}(X)$ for any $X > 0$ and $l = 0, 1, \cdots$. By letting $A \leftarrow \hat{\Theta}_K$, $Y \leftarrow M$, $X \leftarrow \hat{\Phi}$ in \eqref{form111} and using the definition of $\hat{V}_K$, we have
\begin{align}
    &\hspace{-8mm}\underset{\|K'\|_F=1}{\sup} \tr\left(\sum_{l=0}^{\infty}(\hat{\Theta}_K^{\top})^lM\hat{\Theta}_K^l\right) \leq \frac{\|\hat{V}_K\|}{\sigma_{\rm min}(\hat{\Phi})} \underset{\|K'\|_F=1}{\sup} \tr(M) \nonumber \\
    & \leq \frac{Lc}{\rho \sigma_{\rm min}(\hat{\Phi})} \left( 2c / \sigma_{\min }(\hat{\Phi}) + \tr(R)-\tr(\hat{Q})\right) \label{boundofsup2}
\end{align}
where the last inequality follows from \eqref{lm_rho} and \eqref{boundofM}. Finally, by substituting \eqref{boundofgamma}, \eqref{lm_rho}, \eqref{boundofThV}, \eqref{boudnoftrace}, \eqref{new_expa}, and \eqref{boundofsup2} into \eqref{l8_pf}, we have \eqref{def_qqq}. This completes the proof. \qed

\section{Preliminary lemmas for sample complexity analysis}\label{pro_lem_MTgradient L-smooth} 
\begin{lem}
\label{lem_U_ij_stable}
Consider Problem 1 and ${\mathbb K}$ in \eqref{def_Kdom}. Given $K\in{\mathbb K}$, define a set of stabilizing controllers as
\begin{equation}
    \label{stable set}
    {\mathbb S} \coloneqq \{K~|~\hat{\Theta}_{K}~{\rm in}~\eqref{def_THhat}~{\rm is~Schur}\}.
\end{equation}
Then, any $K'$ such that   
\begin{equation}
    \label{stable_condi}
    \|K'- K\| \leq 
    \min \left(
    \frac{\rho\sigma_{\min}(\hat{\Phi})}{4Lch_K}, 2\|K\|
    \right)
\end{equation}
satisfies $K' \in \mathbb S$, where $\hat{\Phi}$ is defined in \eqref{def_Phihat}, $\rho$ is defined in \eqref{def_rhoQ}, and $h_K$ is defined in Lemma \ref{lem_stochastic analysis}. 
\end{lem}
\begin{proof}
The proof consists of the following two steps: 
\begin{itemize}
    \item[i)] First, assuming that $K'$ satisfies both \eqref{stable_condi} and $K' \in \mathbb S$, we show that 
    \begin{eqnarray}
    \hspace{-14mm} \|\hat{V}_{K'} - \hat{V}_{K}\| &\hspace{-2.5mm}\leq\hspace{-2.5mm}& \frac{4L^2c^2(\|\hat{\Theta}_K\| + \|K\|)\|K'-K\|}{\sigma_{\min}(\hat{\Phi})\rho^2} \label{V_K_dif_1}\\
    \hspace{-14mm} \tr(\hat{V}_{K'}) &\hspace{-2.5mm}\leq\hspace{-2.5mm}& \vartheta(K) \coloneqq (Lm +n)\left(\|\hat{V}_{K}\| + \frac{Lc}{\rho}\right)\hspace{-1mm} \label{V_K_dif_1_tr} \\
    \hspace{-14mm} {\rm sprad}(\hat{\Theta}_{K'}) &\hspace{-2.5mm}<\hspace{-2.5mm}& 1-\frac{\sigma_{\min}(\hat{\Phi})}{3\vartheta({K'})}. \label{V_K_dif_1_sprad}
\end{eqnarray}
    \item[ii)]  Second, we show that $K'$ satisfying \eqref{stable_condi} yields $K' \in \mathbb S$ by reductio ad absurdum. We omit this part as it is similar to Lemma 22 in \cite{fazel}.
\end{itemize}
First, we show \eqref{V_K_dif_1}. To this end, assume $K' \in \mathbb S$. Because $K\in {\mathbb S}$, for any $X \geq 0$ there exists 
$\TAU_{K}(X) \coloneqq \sum_{t=0}^{\infty}\hat{\Theta}_{K}^tX(\hat{\Theta}_{K}^{\top})^t$, $\EFF_{K}(X)  \coloneqq \hat{\Theta}_{K}X\hat{\Theta}^{\top}_{K}$. 
We denote the identity operator as ${\rm I}$. In addition, we define the operator norm of $\|\TAU\|$ as $\|\TAU\| \coloneqq \sup_X \|\TAU(X)\| / \|X\|$. To simplify the notation, we define $\mathcal A \coloneqq {\rm I} - \mathcal F_K$ and $\mathcal B \coloneqq \mathcal F_{K'} - \mathcal F_K$. As shown in Lemma 18 of \cite{fazel} and the proof of Lemma 20 in \cite{fazel}, it follows that $\mathcal A^{-1} = ({\rm I} - \mathcal F_K)^{-1} = \TAU_K$. 
Note that $\|P^{\top}\Pi\| \leq 1$, and $\|K' - K\| \leq 2\|K\|$ holds from \eqref{stable_condi}. Hence, similarly to Lemma 19 in \cite{fazel}, we have
\begin{equation}
\|\EFF_{K'}-\EFF_{K}\| \leq 2(\|\hat{\Theta}_{K}\| + \|K\|)\|K' -K\|. \label{bound_EFF}
\end{equation}
Also, from Lemma 17 of \cite{fazel} and \eqref{lm_rho}, we have
\begin{equation}
\label{bound_TAU}
\begin{split}
    \|\TAU_{K}\| \leq \frac{\|\hat{V}_K\|}{\sigma_{\min}(\hat{\Phi})}\leq \frac{Lc}{\rho\sigma_{\min}(\hat{\Phi})}. 
    \end{split}
\end{equation}
Thus, from \eqref{stable_condi}, \eqref{bound_EFF}, and \eqref{bound_TAU}, we have $\|\EFF_{K'}-\EFF_{K}\|\|\TAU_{K}\| \leq 1/2$. Therefore, similarly to Lemma 20 in \cite{fazel}, we have 
\begin{equation}
\label{dif_TAU TAU'}
\begin{split}
    &\|\TAU_{K'}(X) -\TAU_K(X)\| \leq 2\|{\mathcal A}^{-1}\|\|{\mathcal B}\|\|{\mathcal A}^{-1}(X)\|\\
    & \leq \frac{4Lc(\|\hat{\Theta}_K\| + \|K\|)\|{\mathcal A}^{-1}(X)\|\|K'-K\|}{\sigma_{\min}(\hat{\Phi})\rho}. 
    \end{split}
\end{equation}
Taking $X = \hat{\Phi}$ in \eqref{dif_TAU TAU'} and using \eqref{lm_rho}, we have \eqref{V_K_dif_1}. Next, we show \eqref{V_K_dif_1_tr}. Note that 
\begin{align}
\tr(\hat{V}_{K'}) &\leq (Lm +n)\|\hat{V}_{K'}\| \label{trVV} \\ 
&\leq (Lm +n)(\|\hat{V}_{K}\| + \|\hat{V}_{K'}- \hat{V}_{K}\|). \nonumber  
\end{align}
Using \eqref{V_K_dif_1} and \eqref{stable_condi}, we have \eqref{V_K_dif_1_tr}.  
Finally, we show \eqref{V_K_dif_1_sprad}. From Lemma 23 of \cite{fazel}, we have $\tr(\hat{V}_{K'}) \geq  \sigma_{\min}(\hat{\Phi}) / 2(1-\sprad(\hat{\Theta}_{K'}))$. Thus, it follows that 
\begin{equation*}
    \sprad(\hat{\Theta}_{K'}) \leq 1-\frac{\sigma_{\min}(\hat{\Phi})}{2\tr(\hat{V}_{K'})}\leq 1-\frac{\sigma_{\min}(\hat{\Phi})}{2\vartheta({K'})}< 1-\frac{\sigma_{\min}(\hat{\Phi})}{3\vartheta({K'})}
\end{equation*}
this yields \eqref{V_K_dif_1_sprad}. This completes the proof.  
\end{proof}

\begin{lem}
\label{modelfree smooth}
Consider Problem \ref{problem_1} and ${\mathbb K}$ in \eqref{def_Kdom}. Define $\rho$ and $\hat{Q}$ in Lemma \ref{smoothness_new} and $\hat{\Phi}$ in \eqref{def_Phihat}. 
Given $K\in{\mathbb K}$, if $K'$ satisfies
\begin{equation}\label{U_K_condi}
     \|K'-K\| \leq \min \left(
    \frac{c-J(K)}{\epsilon(K)},
   \frac{\rho\sigma_{\min}(\hat{\Phi})}{4Lc(Lm +n)h_K}\hspace{-2pt},
   2\|K\|
    \right) 
\end{equation}
where $\hat{\Theta}_{K}$ is defined in \eqref{def_THhat}, $h_K$ in Lemma~\ref{lem_stochastic analysis}, and $\epsilon$ in \eqref{epsilon}, then $K' \in {\mathbb K} \cap \mathbb U_K$, where $\mathbb U_K$ is defined as
\begin{equation}\label{neighbourhood}
    {\mathbb U}_K \coloneqq
    \{K'~{\rm s.t.}~(1-\tau)K + \tau K'\in {\mathbb K}~{\rm holds}~\forall~\tau \in [0,1]\}.
\end{equation}
 \end{lem}
 \begin{proof}
Note that $K'$ satisfies \eqref{stable_condi}. Hence, $K' \in {\mathbb S}$ follows, where ${\mathbb S}$ in \eqref{stable set}. Because $K\in {\mathbb S}$, for any $X \geq 0$, there exists
$\TAU^{\rm R}_{K}(X)  \coloneqq \sum_{t=0}^{\infty}(\hat{\Theta}_{K}^{\top})^tX\hat{\Theta}_{K}^t$, $\EFF^{\rm R}_{K}(X)  \coloneqq \hat{\Theta}^{\top}_{K}X\hat{\Theta}_{K}$. 
Similarly, $\TAU^{\rm R}_{K'}(X)$ exists. In addition, $J(K) = \tr(\hat{\Phi}\hat{\Psi}_K)$ follows from \eqref{6}. Let $\hat{Q}_K \coloneqq \hat{Q} + \hat{K}^{\top}R\hat{K}$. Then, we have
\begin{align}
    \hspace{-8mm}|J(K') -J(K)| \leq \tr(\hat{\Phi})\|\TAU^{\rm R}_{K'}(\hat{Q}_{K'})-\TAU^{\rm R}_{K}(\hat{Q}_K)\|. \label{JKJKbound}
\end{align}
Further, we have
\begin{equation*}
    \hspace{-8mm}\begin{split}
        &\|\hat{\Psi}_{K'} -\hat{\Psi}_{K}\| = \|\TAU^{\rm R}_{K'}(\hat{Q}_{K'})-\TAU^{\rm R}_{K}(\hat{Q}_K)\|\\
       & =\|\TAU^{\rm R}_{K'}(\hat{Q}_{K'})-\TAU^{\rm R}_{K}(\hat{Q}_{K'}) + \TAU^{\rm R}_{K}(\hat{Q}_{K'}) - \TAU^{\rm R}_{K}(\hat{Q}_K)\|\\
       & = \|(\TAU^{\rm R}_{K'}-\TAU^{\rm R}_{K})\circ \hat{Q}_{K'} 
       + \TAU^{\rm R}_{K}\circ(\hat{Q}_{K'}-\hat{Q}_K)\|\\
       & \leq \|\TAU^{\rm R}_{K'}-\TAU^{\rm R}_{K}\|\| \hat{Q}_{K'}\| + \|\TAU^{\rm R}_K\|\|K'^{\top}RK'- K^{\top}RK\|. 
    \end{split}
\end{equation*}
By substituting this into \eqref{JKJKbound}, we have 
\begin{eqnarray}
    \hspace{-15mm} &&|J(K') -J(K)| \leq  \tr(\hat{\Phi}) \times \nonumber \\
    \hspace{-15mm} &&  \scalebox{0.95}{$\displaystyle  \left(
    \|\TAU^{\rm R}_{K'}-\TAU^{\rm R}_{K}\|\| \hat{Q}_{K'}\| + \|\TAU^{\rm R}_K\|\|K'^{\top}RK'- K^{\top}RK\| \label{cost_dif_2}
    \right). $}
\end{eqnarray}
In what follows, we show upper bounds of the four operator norms in the RHS of \eqref{cost_dif_2}. 

$\bullet$ First, we show
\begin{equation}
    \label{TAU'_bound}
     \|\TAU^{\rm R}_{K}\| 
    \leq \frac{Lc(Lm +n)}{\rho\sigma_{\min}(\hat{\Phi})}.
\end{equation}
Let the unit eigenvector corresponding to the largest eigenvalue of $\TAU^{\rm R}_{K}(X)$ be denoted as $g\in{\mathbb R}^{Lm +n}$. Using \eqref{lm_rho} and the formula 
\begin{align}\label{form123}
 MM^{\top} \leq MXM^{\top} / \sigma_{\rm min}(X)   
\end{align}
for $X > 0$ and any matrix $M$, we have  
\begin{equation*}
    \begin{split}
         & \scalebox{0.94}{$\displaystyle  \|\TAU^{\rm R}_{K}(X)\| = \sum_{t=0}^{\infty}g^{\top}(\hat{\Theta}_{K}^{\top})^tX\hat{\Theta}_K^tg = \sum_{t=0}^{\infty}\tr(g^{\top}(\hat{\Theta}_{K}^{\top})^tX\hat{\Theta}_K^tg)$} \\
         & \leq \|X\|\sum_{t=0}^{\infty}\tr(\hat{\Theta}_K^tgg^{\top}(\hat{\Theta}_K^{\top})^t)\leq\|X\|\sum_{t=0}^{\infty}\tr(\hat{\Theta}_K^t(\hat{\Theta}_K^{\top})^t )\\
         & \leq \frac{\|X\|}{\sigma_{\min}(\hat{\Phi})}\sum_{t=0}^{\infty}\tr(\hat{\Theta}_K^t\hat{\Phi}(\hat{\Theta}_K^t)^{\top}) \leq \frac{\|X\|}{\sigma_{\min}(\hat{\Phi})}\tr(\hat{V}_K)
         \\
         &\leq  \frac{\|X\|Lc(Lm +n)}{\rho\sigma_{\min}(\hat{\Phi})}, 
    \end{split}
\end{equation*}
where the last inequality follows from \eqref{trVV} and \eqref{lm_rho}. This yields \eqref{TAU'_bound}. 

$\bullet$ Second, similarly to deriving the second inequality in \eqref{dif_TAU TAU'}, we have 
\begin{equation}
\label{dif_TAU'}
    \|\TAU^{\rm R}_{K'}- \TAU_{K}^{\rm R}\| \leq \frac{4L^2c^2(Lm +n)^2(\|\hat{\Theta}_K\| + \|K\|)\|K'-K\|}{\rho^2\sigma^2_{\min}(\hat{\Phi})}.
\end{equation}

$\bullet$ Finally, using the relation $K' = K + K' -K$, we have 
\begin{equation*}
   \begin{split}
       & K'^{\top}RK'- K^{\top}RK\\
       &= (K+K' -K)^{\top}R(K+K' -K) - K^{\top}RK \\
       &=  \scalebox{0.92}{$\displaystyle  K^{\top}R(K'-K) + (K'-K)^{\top}RK+(K'-K)^{\top}R(K'-K). $}
   \end{split}
\end{equation*}
Using $\|K'-K\|\leq 2\|K\|$, we have 
\begin{align}
    \|K'^{\top}RK'- K^{\top}RK\|\leq 4\|K\|\|R\|\|K'-K\| \label{K'-K} 
\end{align}
Further, since $\|K'\| \leq \|K\| + \|K'-K\| \leq 3\|K\|$, we have 
\begin{align}
    \|\hat{Q}_{K'}\| \leq  \|\hat{Q}\|+ 9\|K\|^2\|R\|.\label{QK'_bound}
\end{align}
Substituting \eqref{TAU'_bound}, \eqref{dif_TAU'}-\eqref{QK'_bound} into the RHS of \eqref{cost_dif_2}, we have 
\begin{align}\label{lm10_finalsss}
 |J(K')-J(K)|\leq \epsilon(K)\|K'-K\|. 
\end{align}
Since $\|K'-K\| \leq (c-J(K))/\epsilon(K)$ follows from \eqref{U_K_condi}, we have 
 \begin{equation}\label{lm10_final}
    J(K')\leq J(K) + |J(K')-J(K)|  \leq c. 
 \end{equation}
 This shows $K' \in \mathbb K$. Finally, for $K(\tau) \coloneqq (1-\tau)K + \tau K'$ with $\tau \in [0,1]$, we have 
 \[
 \|K(\tau) - K\| = \tau \|K' - K\| \leq  \|K' - K\|. 
 \]
 Hence, the above argument holds even when $K'$ is replaced by $K(\tau)$. Therefore, $K(\tau) \in \mathbb K$ for any $\tau \in [0,1]$. From the definition of $\mathbb U_K$ in \eqref{neighbourhood}, we have $K' \in \mathbb U_K$. This completes the proof. 
 \end{proof}
 
\section{Proof of Lemma~\ref{lem_stochastic analysis}}\label{stochastic analysis}
Note that \eqref{U_K_condi} holds by choosing $K \leftarrow \tilde{K}_i$ and $K' \leftarrow \tilde{K}_{i,j}$ in \eqref{sample policy} because $\|\tilde{K}_{i,j} - \tilde{K}_i\| = \delta \|U_j\| \leq \delta \|U_j\|_F \leq \delta_{\rm st}$. Hence, $\tilde{K}_{i,j} \in \mathbb K$; thus, $\hat{\Theta}_{\tilde{K}_{i,j}}$ is Schur. For any $j\in [0,s-1]$, let $v_{j}(L)\sim{\mathcal N}_{\Phi}$ denote the initial IOH of the $j$-th episode.
Let $\mathcal B$ and $\mathcal S$ denote uniform distributions over $\{X \in \mathbb R^{m \times L(m+r)},~ \|X\|_F \leq 1\}$ and $\{X \in \mathbb R^{m \times L(m+r)},~ \|X\|_F = 1\}$, respectively. Define 
\begin{align}
    & \label{nabla_bar}
    \textstyle \bar{\nabla}J(\tilde{K}_i)  \coloneqq \frac{mL(m+r)}{s\delta}\sum_{j=0}^{s-1}\sum_{t=L}^{\infty}c_{i,j}(t)U_j,\\
    & \label{nabla_N}
    \textstyle \bar{\nabla}_NJ(\tilde{K}_{i})  \coloneqq  \frac{mL(m+r)}{s\delta}\sum_{j=0}^{s-1}\sum_{t=L}^{L+N-1}c_{i,j}(t)U_j,\\
    & \label{nabla_delta}
    \textstyle \nabla_\delta J(\tilde{K}_i)  \coloneqq
    \frac{mL(m+r)}{\delta}{\mathbb E}_{U_j \sim {\mathcal S}}\left[J(\tilde{K}_{i,j})U_j\right ],
\end{align}
where $c_{i,j}(t)$ is defined in \eqref{reward_multiple}. It follows that 
\begin{align}
& \hspace{-8mm}\| \tilde{\nabla}J(\tilde{K}_i)-\nabla J(\tilde{K}_i)\|_F
     \nonumber \\
     &   \hspace{-6mm} \leq \left\|
     \tilde{\nabla}J(\tilde{K}_i)- 
     {\mathbb E}_{U_j\sim{\mathcal S}}\left[
     {\mathbb E}_{v_j(L)\sim {\mathcal N}_{\Phi}}\left[
     {\bar{\nabla}}_NJ(\tilde{K}_i)
          \right]
     \right]
     \right\|_F 
      \nonumber \\  
       & \hspace{-4mm}+ \left\|
       {\mathbb E}_{U_j\sim{\mathcal S}}\left[
     {\mathbb E}_{v_j(L)\sim {\mathcal N}_{\Phi}}\left[
     {\bar{\nabla}}_N J(\tilde{K}_i)
          \right]
     \right] - \nabla_{\delta} J(\tilde{K}_i)
       \right\|_F \nonumber \\
       & \hspace{-4mm}+\|\nabla_{\delta} J(\tilde{K}_i) - \nabla J(\tilde{K}_i)\|_F. \label{grad_dif}
\end{align}
We derive upper bounds of the three terms in the RHS of \eqref{grad_dif} as follows.

$\bullet$~First, we show that a bound of the second term is given as 
\begin{equation}\label{lm11_pf1}
\left\| {\mathbb E}_{U_j\sim{\mathcal S}}
        \left[
        {\mathbb E}_{v_j(L)\sim{\mathcal N}_{\Phi}}
    \left[
    \bar{\nabla}_NJ(\tilde{K}_i)
    \right]
        \right] - \nabla_{\delta}J(\tilde{K}_i)
        \right\|_F \leq \chi(\tilde{K}_i). 
\end{equation}
Let $J_{N}(K)  \coloneqq {\mathbb E}\left[
   \sum_{t=L}^{L +N-1}y^{\top}(t)Qy(t) + u^{\top}(t)Ru(t)
   \right]$. For showing \eqref{lm11_pf1}, we show 
\begin{equation}\label{cost_dif_inf_fin}
    \ J(K)-J_N(K) \leq  \frac{(Lm+n)L^2c^2\|\hat{Q}_K\|}{N\sigma_{\min}(\hat{\Phi})\rho^2}
\end{equation}
for any $K\in {\mathbb K}$, where $\hat{Q}_K\coloneqq \hat{Q} + \hat{K}^{\top}R\hat{K}$. 
Let $ \hat{v}_K(t) \coloneqq \hat{\Theta}_K^{t-L}\hat{v}(L)$ and $\hat{V}_{K,N}\coloneqq {\mathbb E}_{\hat{v}(L)\sim {\mathcal N}_{\hat{\Phi}}}\left[\sum_{t=L}^{N+L-1}\hat{v}_K(t)\hat{v}^{\top}_K(t)\right]$. 
It holds that 
\begin{equation*}
    \begin{split}
    &J(K)-J_N(K) = \tr(\hat{V}_K\hat{Q}_K-\hat{V}_{K,N}\hat{Q}_K)\\
       & \leq \|\hat{Q}_K\|\tr(\hat{V}_K-\hat{V}_{K,N})\\
        &= \|\hat{Q}_K\|\tr\left({\mathbb E}_{\hat{v}(L)\sim{\mathcal N}_{\hat{\Phi}}}\left[\textstyle \sum_{t=L+N}^{\infty}\hat{v}_{K}(t)\hat{v}_K^{\top}(t)\right]\right)\\
       &= \|\hat{Q}_K\|\tr(\hat{\Theta}_{K}^N\hat{V}_{K}(\hat{\Theta}_K^{\top})^N) \\
       &\leq \|\hat{Q}_K\| \|\hat{V}_K\|\tr(\hat{\Theta}_K^N(\hat{\Theta}_K^{\top})^N)
        \hspace{-0.5mm}\leq  \hspace{-0.5mm}  \scalebox{0.95}{$\displaystyle  \frac{\|\hat{Q}_K\|\|\hat{V}_K\|\tr(\hat{\Theta}_K^N\hat{\Phi}(\hat{\Theta}_K^{\top})^N)}{\sigma_{\min}(\hat{\Phi})} $}
    \end{split}
\end{equation*}
where the last inequality follows by using the formula \eqref{form123}. Further, since $\hat{\Theta}_K$ is Schur, we have 
\begin{eqnarray*}
N\tr(\hat{\Theta}_K^N\hat{\Phi}(\hat{\Theta}_K^{\top})^N) 
\leq \sum_{\tau=0}^{N-1}\tr(\hat{\Theta}_K^\tau\hat{\Phi}(\hat{\Theta}_K^{\top})^\tau) \leq \tr(\hat{V}_K).  
\end{eqnarray*}
Hence, $\tr(\hat{\Theta}_K^N\hat{\Phi}(\hat{\Theta}_K^{\top})^N)\leq \tr(\hat{V}_K)/N$ follows. Thus, we have $J - J_N \leq (\|\hat{Q}_K\| \|\hat{V}_K\| / \sigma_{\rm min}(\hat{\Phi})) (\tr (\hat{V}_K)/N)$. By substituting the fact that $\tr(\hat{V}_K) \leq (Lm+n)\|\hat{V}_K\|$ and \eqref{lm_rho} into this relation, we have \eqref{cost_dif_inf_fin}. Note here that the formula
\begin{align}
    \tr(Y MM^{\top}) \leq \tr \left(Y MXM^{\top} / \sigma_{\rm min}(X)\right)
\end{align}
follows for any $X > 0$, $Y \geq 0$, and matrix $M$ from \eqref{form123}. Using this with $Y \leftarrow \hat{Q}_{\tilde{K}_{i,j}}$, $M \leftarrow \hat{\Theta}_{\tilde{K}_{i,j}}$, $X \leftarrow \hat{\Phi}$, we have
\begin{align*}
    &\hspace{-9mm}\textstyle \sum_{t=L}^{\infty}c_{i,j}(t) - \sum_{t=L}^{L+N-1}c_{i,j}(t)\\
    &\hspace{-9mm}\leq \|v_j(L)\|^2 \tr\left( \textstyle \sum_{t=N}^{\infty}(\hat{\Theta}^{\top}_{\tilde{K}_{i,j}})^t\hat{Q}_{\tilde{K}_{i,j}}\hat{\Theta}_{\tilde{K}_{i,j}}^t  \right)\\
    &\hspace{-9mm}\leq \frac{v_{\max}^2}{\sigma_{\min}(\hat{\Phi})}\left(
        J(\tilde{K}_{i,j}) - J_N(\tilde{K}_{i,j})
        \right)
        \\
    &\hspace{-9mm}\leq \left((Lm+n)L^2c^2\|\hat{Q}_{\tilde{K}_{i,j}}\|v_{\max}^2 \right)/\left(N\sigma^2_{\min}(\hat{\Phi})\rho^2 \right)\\
    &\hspace{-9mm}  \scalebox{0.95}{$\displaystyle \leq \left((m+r)L^3c^2(\|\hat{Q}\|+ 9\|\tilde{K}_{i}\|^2\|R\|)v_{\max}^2 \right) / \left(N\sigma^2_{\min}(\hat{\Phi})\rho^2 \right) $}
\end{align*}
where the third inequality follows from \eqref{cost_dif_inf_fin}, and the last one follows from \eqref{QK'_bound} with $K' \leftarrow \tilde{K}_{i,j}$ and $K \leftarrow \tilde{K}_i$, and $n \leq Lr$ because \eqref{condition L} holds. Thus, it follows that 
\begin{equation*}
    \begin{split}
        & \left\| {\mathbb E}_{U_j\sim{\mathcal S}}
        \left[
        {\mathbb E}_{v_j(L)\sim{\mathcal N}_{\Phi}}
    \left[
    \bar{\nabla}J_N(\tilde{K}_i)
    \right]
        \right] - \nabla_{\delta}J(\tilde{K}_i)
        \right\|_F\\
        &  \scalebox{0.72}{$\displaystyle =\frac{mL(m+r)}{s\delta}\left\|
        {\mathbb E}_{U_j\sim{\mathcal S}}\left[
        {\mathbb E}_{v_j(L) \sim {\mathcal N}_{\Phi}}
        \sum_{j=0}^{s-1}\left[
        \sum_{t=L}^{\infty}c_{i,j}(t)U_j - \sum_{t=L}^{L +N-1}c_{i,j}(t)U_j
        \right]
        \right]
        \right\|_F$}\\
        & 
        \scalebox{0.68}{$\displaystyle \leq \frac{mL(m+r)}{s\delta}
        {\mathbb E}_{U_j\sim{\mathcal S}}\left[
        {\mathbb E}_{v_j(L) \sim {\mathcal N}_{\Phi}}
        \left[\left\|\sum_{j=0}^{s-1}\left(
        \sum_{t=L}^{\infty}c_{i,j}(t)U_j - \sum_{t=L}^{L +N-1}c_{i,j}(t)U_j\right)\right\|_F
        \right]
        \right]
        $}\\
       & 
        \scalebox{0.8}{$\displaystyle \leq\frac{mL(m+r)}{s\delta}
        {\mathbb E}_{U_j\sim{\mathcal S}}\left[
        {\mathbb E}_{v_j(L) \sim {\mathcal N}_{\Phi}}
        \sum_{j=0}^{s-1}\left[
        \sum_{t=L}^{\infty}c_{i,j}(t)- \sum_{t=L}^{L +N-1}c_{i,j}(t)
        \right]
        \right]
        $}\\
      &  \scalebox{1.0}{$\displaystyle \leq \frac{m(m+r)^2L^4c^2(\|\hat{Q}\|+ 9\|\tilde{K}_{i}\|^2\|R\|)v_{\max}^2}{N\delta \sigma^2_{\min}(\hat{\Phi})\rho^2} = \chi(\tilde{K}_i)$},
    \end{split}
\end{equation*}
where the second inequality follows from Jensen's inequality. This shows \eqref{lm11_pf1}. 


$\bullet$~Second, we show
\begin{equation}\label{lm11_pf2}
    \|\nabla_{\delta}J(\tilde{K}_{i})-\nabla J(\tilde{K}_i)\|_F \leq q\delta. 
\end{equation}
Note from Lemma 1 of \cite{flaxman2004online} that 
${\mathbb E}_{W \sim {\mathcal B}}\left[\nabla J(\tilde{K}_{i} +\delta W)\right] = \nabla_{\delta}J(\tilde{K}_i)$ holds. Because $\delta \leq \delta_{\rm st}$, $\tilde{K}_i + \delta W \in {\mathbb U}_{\tilde{K}_i}$ holds from Lemma \ref{modelfree smooth}. Hence, \eqref{def_qqq} holds for $K' \leftarrow \tilde{K}_i+\delta W$ and $\tilde{K}_i \leftarrow K$, which is equivalent to $\|\nabla J(\tilde{K}_i +\delta W)- \nabla J(\tilde{K}_i)\|_F \leq q \|\delta W\|_F \leq q\delta$. Therefore, we have  
\begin{equation*}
\begin{split}
& {\rm LHS~of~(\ref{lm11_pf2})} = \left\|
{\mathbb E}_{W \sim {\mathcal B}}\left[
\nabla J(\tilde{K}_i +\delta W)- \nabla J(\tilde{K}_i)
\right]
\right\|_F\\
&\leq {\mathbb E}_{W \sim {\mathcal B}}\left[
\left\|\nabla J(\tilde{K}_i +\delta W) -\nabla J(\tilde{K}_i)
\right\|_F  
\right]\leq q\delta,
\end{split}
\end{equation*}
showing \eqref{lm11_pf2}, where the first inequality follows from Jensen's inequality. 

$\bullet$~Third, we show
    \begin{align}
    &\hspace{-8mm} {\mathbb P}\left(
    \left\|
     \tilde{\nabla}J(\tilde{K}_i)- 
     {\mathbb E}_{U_j\sim{\mathcal S}}\left[
     {\mathbb E}_{v_j(L)\sim {\mathcal N}_{\Phi}}\left[
     \bar{\nabla}_NJ(\tilde{K}_i)
          \right]
     \right]
     \right\|_F \leq o
    \right) \nonumber \\
    &\hspace{45mm} \geq {\rm Pr}(\tilde{K}_i) \label{bern}
    \end{align}
    holds for any $o > 0$. This follows from Corollary 6.2.1 of \cite{tropp2014introduction}, with the replacement of 
    $B$ in the corollary as $\mathbb E[\mathbb E[\bar{\nabla}_NJ]]$, 
    $R$ as $\tilde{\nabla}J$, 
    $L$ as $\zeta(\tilde{K}_i)/\sqrt{m}$, 
    $m_2$ as $\zeta^2(\tilde{K}_i)/m$, and
    $t$ as $o/\sqrt{m}$. The first two replacements are obvious. The last follows from the fact that for $M \in \mathbb R^{d_1 \times d_2}$ such that $d_1 \leq d_2$, $\|M\|_F \leq o$ holds if $\|M\| \leq o/\sqrt{d_1}$. We show the third and fourth replacements are possible. Let 
    \begin{equation*}
    X_{j} \coloneqq \textstyle \frac{mL(m+r)}{\delta} \sum_{t=L}^{L +N-1}c_{i,j}(t)U_j.
\end{equation*}
    Note that \eqref{lm10_finalsss}-\eqref{lm10_final} holds when $K \leftarrow \tilde{K}_i$ and $K' \leftarrow \tilde{K}_{i,j}$. Hence, 
    \[
    J(\tilde{K}_{i,j}) \leq J(\tilde{K}_i) + \epsilon(\tilde{K}_i) \delta \leq 2 J(\tilde{K}_i)
    \]
    holds because $\delta \leq \delta_{\rm st}$ in \eqref{delta_st}. Hence, using the fact that $\|U_j\| \leq 1$ if $\|U_j\|_F \leq 1$, we have
\begin{align*}
       & \hspace{-6mm}\|X_{j}\| \leq \frac{mL(m+r)}{\delta}\sum_{t=L}^{\infty}c_{i,j}(t)\\
        & \hspace{-6mm}=  \frac{mL(m+r)}{\delta} \sum_{t=0}^{\infty}\tr\left(
        \hat{v}^{\top}_j(L)(\hat{\Theta}_{\tilde{K}_{i,j}}^{\top})^t\hat{Q}_{\tilde{K}_{i,j}}\hat{\Theta}_{\tilde{K}_{i,j}}^t\hat{v}_j(L)
        \right)\\
         &\hspace{-6mm} \leq  \frac{mL(m+r)v_{\max}^2}{\delta\sigma_{\min}(\hat{\Phi})} \sum_{t=0}^{\infty}\tr\left(
       (\hat{\Theta}_{\tilde{K}_{i,j}}^{\top})^t\hat{Q}_{\tilde{K}_{i,j}}\hat{\Theta}_{\tilde{K}_{i,j}}^t\hat{\Phi} \right)\\
       &\hspace{-6mm} =  \frac{mL(m+r)v_{\max}^2J(\tilde{K}_{i,j})}{\delta\sigma_{\min}(\hat{\Phi})} \leq  \frac{2mL(m+r)v_{\max}^2J(\tilde{K}_{i})}{\delta\sigma_{\min}(\hat{\Phi})}. 
\end{align*}
Thus, $\|X_{j}\| \leq \zeta(\tilde{K}_i) / \sqrt{m}$. Note that this inequality holds for any $j$. Thus, we can choose $\zeta(\tilde{K}_i)/\sqrt{m}$ as $L$ in the corollary. Also, since $\|\mathbb E[X_jX_j^{\top}]\| \leq \mathbb E[\|X_jX_j^{\top}\|]$, we can see from a simple calculation that $\zeta^2(\tilde{K}_i)/m$ can be chosen as $m_2$ in the corollary. 

$\bullet$ Finally, by substituting \eqref{lm11_pf1}, \eqref{lm11_pf2}, and \eqref{bern} into \eqref{grad_dif}, we have \eqref{stoc_eq}. This completes the proof.
\qed{}

\section{Proof of Theorem~\ref{linear convergent some trajectory}}\label{linear conv modelfree}
First, we show Claim i). Note that $g_0(o) > 0$ for any $o>0$ and that $\check{K}^{[\alpha]}$, $h_{\check{K}^{[\alpha]}}$, and $\epsilon(\check{K}^{[\alpha]})$ are continuous around a positive neighborhood at $\alpha = 0$. Hence, Claim i) follows. Next, we show 
\begin{align}
    &\hspace{-8mm}{\mathbb P}\left(
    J(\tilde{K}_{i+1}) - J(K^{\star})  \leq 
    \tilde{\beta}(\tilde{K}_i)((J(\tilde{K}_i)-J(K^{\star})) \right) \nonumber \\
    &\hspace{45mm} \geq {\rm Pr}(\tilde{K}_i, o) \label{theo2_eq_pf}
\end{align}
for any $o, \alpha >0$ such that $\alpha < g_{\alpha}(o)$. To this end, we show 
\begin{equation}\label{thm3_stab}
    \mathbb P(\tilde{K}_{i+1}\in {\mathbb S}) \geq {\rm Pr}(\tilde{K}_i), 
\end{equation}
where $\mathbb S$ is defined in \eqref{stable set}. From Theorem \ref{linear_convergence}, we have $J(\check{K}^{[\alpha]}) - J(K^{\star}) \leq \beta(\tilde{K}_i) (J(\tilde{K}_i)-J(K^{\star}))$ with $\beta < 1$, which yields $\check{K}^{[\alpha]} \in {\mathbb K}$. Since $\|X\| \leq \|X\|_F$ for any $X$ and \eqref{stoc_eq} follows, we have 
\begin{equation}
\|\tilde{K}_{i+1} - \check{K}^{[\alpha]}\| \leq \alpha \|\tilde{\nabla} J(\tilde{K}_i) -\nabla J(\tilde{K}_i)\|_F \leq \alpha \theta_o \label{thm2Prrr}
\end{equation}
with a probability of at least ${\rm Pr}$. Based on this, by letting $K \leftarrow \check{K}^{[\alpha]}$ and $K' \leftarrow \tilde{K}_{i+1}$ in Lemma \ref{lem_U_ij_stable}, the relation \eqref{stable_condi} holds with the same probability because $\alpha < g_{\alpha}(o)$ holds. Therefore, \eqref{thm3_stab} holds. Now, we prove \eqref{theo2_eq_pf}. Let $\tilde{K}'' \coloneqq \tilde{K}_{i}-(R + \Pi^{\top}\Psi_{\tilde{K}_i}\Pi)^{-1}E_{\tilde{K}_i}$. 
From the proof of Lemma 11 in \cite{fazel}, it follows 
\begin{equation*}
J(\tilde{K}_i)-J(K^{\star}) \geq \tr(\hat{V}_{\tilde{K}''}\hat{E}_{\tilde{K}_i}^{\top}(R + \hat{\Pi}^{\top}\hat{\Psi}_{\tilde{K}_i}\hat{\Pi})^{-1}\hat{E}_{\tilde{K}_i})
\end{equation*}
where $\hat{V}_K$ is defined in Lemma~\ref{PL Lem MIMO}, $\hat{E}_K$ in \eqref{re gradient}, and $\hat{\Pi} \coloneqq P^{\top}\Pi$. Note here that $\hat{V}_{\tilde{K}''} \geq \hat{\Phi}$. Hence, 
\begin{align}
    \hspace{-8mm} J(\tilde{K}_i)-J(K^{\star}) &\geq \tr(P\hat{\Phi}P^{\top}E_{\tilde{K}_i}^{\top}(R + \Pi^{\top}\Psi_{\tilde{K}_i}\Pi)^{-1}E_{\tilde{K}_i}) \nonumber \\
    &\geq \epsilon^*(\tilde{K}_i).  \label{HHJ}
\end{align}
Further, from \eqref{thm2Prrr} and $\alpha < g_{\alpha}(o)$, \eqref{U_K_condi} holds when $K \leftarrow \check{K}^{[\alpha]}$ and $K' \leftarrow \tilde{K}_{i+1}$. Hence, from \eqref{lm10_finalsss} and \eqref{thm2Prrr}, the relation $|J(\tilde{K}_{i+1}) - J(\check{K}^{[\alpha]})| \leq \epsilon(\check{K}^{[\alpha]}) \alpha \theta_o$ holds with a probability of at least ${\rm Pr}$. Hence, from $\alpha < g_{\alpha}(o)$ and \eqref{HHJ}, it follows that 
\begin{equation}
\label{L_eq_1}
         |J(\tilde{K}_{i+1})-J(\check{K}^{[\alpha]})|\leq \frac{1-\beta(\tilde{K}_i)}{2}\left(J(\tilde{K}_i) - J(K^{\star})\right)
\end{equation}
with a probability of at least ${\rm Pr}$. Furthermore, it follows that
\begin{equation}\label{lin_conv_thm3}
    J(\check{K}^{[\alpha]}) - J(K^{\star}) \leq \beta(\tilde{K}_i) (J(\tilde{K}_i)-J(K^{\star})). 
\end{equation}
Note that $a+b \leq |a|+b \leq c+d$ if $|a|\leq c$ and $b \leq d$. By adding \eqref{lin_conv_thm3} to \eqref{L_eq_1}, \eqref{theo2_eq_pf} holds for any $o > 0$. Finally, by letting $o$ be $\bar{o}$, we have \eqref{theo2_eq_1}. This completes the proof. \qed{}

\printcredits

\end{document}